\documentclass[reprint]{revtex4-1}

\usepackage[usenames,dvipsnames]{color}
\usepackage{graphicx,epstopdf}
\usepackage[bookmarks]{hyperref}
\usepackage[normalem]{ulem}

\usepackage{comment}

\usepackage{amsmath}
\usepackage{amsfonts}
\usepackage{amssymb}

\usepackage{multirow}

\usepackage{lineno}
\newcommand{\result}[1]{#1}

\newcommand{\externalresult}[1]{#1}

\newcommand{\LandryEssick}{LE}

\newcommand{\GP}{GP}
\newcommand{\GPs}{{\GP}s}
\newcommand{\EOS}{EOS}
\newcommand{\EOSs}{{\EOS}s}
\newcommand{\GW}{{GW}}
\newcommand{\GWs}{{\GW}s}
\newcommand{\NS}{NS}
\newcommand{\NSs}{{\NS}s}
\newcommand{\BH}{BH}
\newcommand{\BHs}{{\BH}s}

\newcommand{\BNS}{\ensuremath{\mathrm{BNS}}}
\newcommand{\BBH}{\ensuremath{\mathrm{BBH}}}
\newcommand{\NSBH}{\ensuremath{\mathrm{NSBH}}}
\newcommand{\BHNS}{\ensuremath{\mathrm{BHNS}}}

\newcommand{\KDE}{KDE}

\newcommand{\rhonuc}{\ensuremath{\rho_\mathrm{nuc}}}
\newcommand{\Mmax}{\ensuremath{M_\mathrm{max}}}
\newcommand{\Fbrk}{\ensuremath{f_\mathrm{max}}}
\newcommand{\Chibrk}{\ensuremath{\chi_\mathrm{max}}}
\newcommand{\Fdyn}{\ensuremath{f_\mathrm{dyn}}}

\newcommand{\Meject}{\result{M_\mathrm{ejecta}^{(\mathrm{dyn})}}}
\newcommand{\Veject}{\result{v_\mathrm{ejecta}^{(\mathrm{dyn})}/c}}

\newcommand{\MejectAgnostic}{\result{\ensuremath{6.5^{+6.3}_{-3.9}\times10^{-3}}}}
\newcommand{\MejectInformed}{\result{\ensuremath{3.9^{+2.4}_{-1.3}\times10^{-3}}}}
\newcommand{\VejectAgnostic}{\result{\ensuremath{0.257^{+0.027}_{-0.034}}}}
\newcommand{\VejectInformed}{\result{\ensuremath{0.231^{+0.014}_{-0.014}}}}

\newcommand{\MOneInformed}{\result{\ensuremath{1.46^{+0.11}_{-0.10}}}}
\newcommand{\MTwoInformed}{\result{\ensuremath{1.28^{+0.08}_{-0.09}}}}
\newcommand{\LOneInformed}{\result{\ensuremath{380^{+249}_{-231}}}}
\newcommand{\LTwoInformed}{\result{\ensuremath{844^{+553}_{-405}}}}
\newcommand{\LTildeInformed}{\result{\ensuremath{572^{+254}_{-212}}}}
\newcommand{\ROneInformed}{\result{\ensuremath{12.50^{+0.98}_{-0.87}}}}
\newcommand{\RTwoInformed}{\result{\ensuremath{12.51^{+1.02}_{-0.96}}}}
\newcommand{\RhocOneInformed}{\result{\ensuremath{7.43^{+1.48}_{-1.23}}}}
\newcommand{\RhocTwoInformed}{\result{\ensuremath{6.65^{+1.03}_{-1.04}}}}

\newcommand{\MOneAgnostic}{\result{\ensuremath{1.49^{+0.13}_{-0.13}}}}
\newcommand{\MTwoAgnostic}{\result{\ensuremath{1.25^{+0.10}_{-0.10}}}}
\newcommand{\LOneAgnostic}{\result{\ensuremath{148^{+274}_{-125}}}}
\newcommand{\LTwoAgnostic}{\result{\ensuremath{430^{+519}_{-301}}}}
\newcommand{\LTildeAgnostic}{\result{\ensuremath{245^{+361}_{-160}}}}
\newcommand{\ROneAgnostic}{\result{\ensuremath{10.88^{+1.99}_{-1.37}}}}
\newcommand{\RTwoAgnostic}{\result{\ensuremath{10.82^{+2.14}_{-1.55}}}}
\newcommand{\RhocOneAgnostic}{\result{\ensuremath{9.79^{+3.21}_{-3.72}}}}
\newcommand{\RhocTwoAgnostic}{\result{\ensuremath{8.85^{+2.52}_{-3.18}}}}

\newcommand{\CanonicalLAgnostic}{\result{\ensuremath{211^{+312}_{-137}}}}
\newcommand{\CanonicalRAgnostic}{\result{\ensuremath{10.86^{+2.04}_{-1.42}}}}
\newcommand{\CanonicalIAgnostic}{\result{\ensuremath{1.25^{+0.35}_{-0.22}}}}
\newcommand{\CanonicalMbAgnostic}{\result{\ensuremath{1.591^{+0.049}_{-0.053}}}}
\newcommand{\MmaxAgnostic}{\result{\ensuremath{2.064^{+0.260}_{-0.134}}}}

\newcommand{\CanonicalLInformed}{\result{\ensuremath{491^{+216}_{-181}}}}
\newcommand{\CanonicalRInformed}{\result{\ensuremath{12.51^{+1.00}_{-0.88}}}}
\newcommand{\CanonicalIInformed}{\result{\ensuremath{1.55^{+0.17}_{-0.16}}}}
\newcommand{\CanonicalMbInformed}{\result{\ensuremath{1.566^{+0.025}_{-0.021}}}}
\newcommand{\MmaxInformed}{\result{\ensuremath{2.017^{+0.238}_{-0.087}}}}

\newcommand{\CanonicalRhocInformed}{\result{\ensuremath{7.15^{+1.12}_{-1.04}\times10^{14}}}}
\newcommand{\POneRhonucInformed}{\result{\ensuremath{4.25^{+0.76}_{-2.10}\times10^{33}}}}
\newcommand{\PTwoRhonucInformed}{\result{\ensuremath{4.44^{+1.09}_{-1.34}\times10^{34}}}}
\newcommand{\PSixRhonucInformed}{\result{\ensuremath{6.84^{+5.58}_{-1.12}\times10^{35}}}}

\newcommand{\CanonicalRhocAgnostic}{\result{\ensuremath{9.40^{+2.78}_{-3.41}\times10^{14}}}}
\newcommand{\POneRhonucAgnostic}{\result{\ensuremath{2.26^{+4.01}_{-2.14}\times10^{33}}}}
\newcommand{\PTwoRhonucAgnostic}{\result{\ensuremath{1.81^{+2.80}_{-1.80}\times10^{34}}}}
\newcommand{\PSixRhonucAgnostic}{\result{\ensuremath{8.56^{+4.81}_{-3.88}\times10^{35}}}}

\newcommand{\PBNSLoSpinInformed}{\result{\ensuremath{ (14.3 \pm 4.5) \%}}} % 0.0978824960911 vs 0.187255874778
\newcommand{\PBHNSLoSpinInformed}{\result{\ensuremath{(23.6 \pm 0.5) \%}}} % 0.240808411782 vs 0.230375505287
\newcommand{\PNSBHLoSpinInformed}{\result{\ensuremath{(54.4 \pm 1.2) \%}}} % 0.556180361032 vs 0.532081560243
\newcommand{\PBBHLoSpinInformed}{\result{\ensuremath{ ( 7.8 \pm 2.7) \%}}} % 0.105128731095 vs 0.0502870596925

\newcommand{\PBNSLoSpinAgnostic}{\result{\ensuremath{ (25.9 \pm 7.1) \%}}} % 0.188294285592 vs 0.330644435298
\newcommand{\PBHNSLoSpinAgnostic}{\result{\ensuremath{(27.6 \pm 1.8) \%}}} % 0.293368349129 vs 0.258130681048
\newcommand{\PNSBHLoSpinAgnostic}{\result{\ensuremath{(38.3 \pm 2.2) \%}}} % 0.405683551244 vs 0.361012491134
\newcommand{\PBBHLoSpinAgnostic}{\result{\ensuremath{ ( 8.1 \pm 3.1) \%}}} % 0.112653814035 vs 0.0502123925203

\newcommand{\PBNSHiSpinInformed}{\result{\ensuremath{ (11.2 \pm 3.7) \%}}} % 0.0751853350924 vs 0.148842262158
\newcommand{\PBHNSHiSpinInformed}{\result{\ensuremath{(18.1 \pm 0.1) \%}}} % 0.181912918226 vs 0.180067835017
\newcommand{\PNSBHHiSpinInformed}{\result{\ensuremath{(61.0 \pm 0.3) \%}}} % 0.61303244235 vs 0.606813854227
\newcommand{\PBBHHiSpinInformed}{\result{\ensuremath{ ( 9.7 \pm 3.3) \%}}} % 0.129869304332 vs 0.064276048598

\newcommand{\PBNSHiSpinAgnostic}{\result{\ensuremath{ (23.9 \pm 6.9) \%}}} % 0.169587414115 vs 0.308531763316
\newcommand{\PBHNSHiSpinAgnostic}{\result{\ensuremath{(25.2 \pm 1.4) \%}}} % 0.265980258303 vs 0.238034154446
\newcommand{\PNSBHHiSpinAgnostic}{\result{\ensuremath{(40.4 \pm 1.7) \%}}} % 0.420684618737 vs 0.387168885408
\newcommand{\PBBHHiSpinAgnostic}{\result{\ensuremath{ (10.5 \pm 3.9) \%}}} % 0.143747708845 vs 0.066265196830

\newcommand{\BayesNSBHHispinAgnostic}{\result{\ensuremath{3.3 \pm 1.4}}} % Bayes factor with P(BNS)=P(NSBH)=P(BHNS)=1/6 and P(BBH)=1/2
\newcommand{\lnBayesBNSBBHHispinAgnostic}{\result{\ensuremath{0.85 \pm 0.69}}} % Bayes factor between BNS and BBH
\newcommand{\BayesNSBHBNSAgnostic}{\result{\ensuremath{1.87 \pm 0.61}}} % Bayes factor between NSBH and BNS
\newcommand{\PHadInformed}{\result{\ensuremath{28\%}}}
\newcommand{\PHypInformed}{\result{\ensuremath{16\%}}}
\newcommand{\PQrkInformed}{\result{\ensuremath{56\%}}}

\newcommand{\PHadAgnostic}{\result{\ensuremath{50\%}}}
\newcommand{\PHypAgnostic}{\result{\ensuremath{14\%}}}
\newcommand{\PQrkAgnostic}{\result{\ensuremath{36\%}}}

\newcommand{\BayesBranchesMrgAgnostic}{\result{\ensuremath{2.0513 \pm (9.5\times10^{-3})}}}
\newcommand{\BayesBranchesMrgInformed}{\result{\ensuremath{\text{unresolved}}}}
\newcommand{\BayesBranchesApprox}{\result{\ensuremath{2}}}

\newcommand{\BayesBranchesHadAgnostic}{\result{\ensuremath{1.5274 \pm (6.8\times10^{-3})}}}
\newcommand{\BayesBranchesHypAgnostic}{\result{\ensuremath{4.084 \pm (1.9\times10^{-2})}}}
\newcommand{\BayesBranchesQrkAgnostic}{\result{\ensuremath{1.684 \pm (1.7\times10^{-2})}}}

\newcommand{\BayesAgnInf}{\result{\ensuremath{1.7952 \pm (1.1\times10^{-3})}}}

\newcommand{\CanonicalLHadAgn}{\result{\ensuremath{185^{+228}_{-119}}}}
\newcommand{\CanonicalRHadAgn}{\result{\ensuremath{10.65^{+1.69}_{-1.42}}}}
\newcommand{\CanonicalIHadAgn}{\result{\ensuremath{1.21^{+0.29}_{-0.21}}}}
\newcommand{\CanonicalMbHadAgn}{\result{\ensuremath{1.591^{+0.055}_{-0.057}}}}
\newcommand{\MmaxHadAgn}{\result{\ensuremath{2.091^{+0.272}_{-0.161}}}}

\newcommand{\CanonicalRhocHadAgn}{\result{\ensuremath{9.95^{+2.96}_{-3.36}\times10^{14}}}}
\newcommand{\POneRhonucHadAgn}{\result{\ensuremath{2.09^{+4.25}_{-1.97}\times10^{33}}}}
\newcommand{\PTwoRhonucHadAgn}{\result{\ensuremath{1.54^{+2.21}_{-1.53}\times10^{34}}}}
\newcommand{\PSixRhonucHadAgn}{\result{\ensuremath{9.50^{+5.12}_{-3.45}\times10^{35}}}}

\newcommand{\CanonicalLHypAgn}{\result{\ensuremath{399^{+296}_{-285}}}}
\newcommand{\CanonicalRHypAgn}{\result{\ensuremath{12.15^{+1.37}_{-1.99}}}}
\newcommand{\CanonicalIHypAgn}{\result{\ensuremath{1.47^{+0.26}_{-0.34}}}}
\newcommand{\CanonicalMbHypAgn}{\result{\ensuremath{1.602^{+0.048}_{-0.024}}}}
\newcommand{\MmaxHypAgn}{\result{\ensuremath{2.038^{+0.225}_{-0.108}}}}

\newcommand{\CanonicalRhocHypAgn}{\result{\ensuremath{7.89^{+3.27}_{-2.17}\times10^{14}}}}
\newcommand{\POneRhonucHypAgn}{\result{\ensuremath{4.22^{+2.80}_{-3.71}\times10^{33}}}}
\newcommand{\PTwoRhonucHypAgn}{\result{\ensuremath{3.34^{+2.09}_{-3.10}\times10^{34}}}}
\newcommand{\PSixRhonucHypAgn}{\result{\ensuremath{7.18^{+3.43}_{-2.63}\times10^{35}}}}

\newcommand{\CanonicalLQrkAgn}{\result{\ensuremath{228^{+291}_{-140}}}}
\newcommand{\CanonicalRQrkAgn}{\result{\ensuremath{10.91^{+1.83}_{-1.31}}}}
\newcommand{\CanonicalIQrkAgn}{\result{\ensuremath{1.28^{+0.33}_{-0.20}}}}
\newcommand{\CanonicalMbQrkAgn}{\result{\ensuremath{1.588^{+0.046}_{-0.052}}}}
\newcommand{\MmaxQrkAgn}{\result{\ensuremath{2.042^{+0.242}_{-0.112}}}}

\newcommand{\CanonicalRhocQrkAgn}{\result{\ensuremath{9.07^{+2.73}_{-3.09}\times10^{14}}}}
\newcommand{\POneRhonucQrkAgn}{\result{\ensuremath{1.91^{+3.58}_{-1.78}\times10^{33}}}}
\newcommand{\PTwoRhonucQrkAgn}{\result{\ensuremath{1.87^{+2.87}_{-1.85}\times10^{34}}}}
\newcommand{\PSixRhonucQrkAgn}{\result{\ensuremath{7.84^{+3.93}_{-3.59}\times10^{35}}}}

\newcommand{\CanonicalLHadInf}{\result{\ensuremath{561^{+163}_{-137}}}}
\newcommand{\CanonicalRHadInf}{\result{\ensuremath{12.98^{+0.55}_{-0.56}}}}
\newcommand{\CanonicalIHadInf}{\result{\ensuremath{1.61^{+0.12}_{-0.11}}}}
\newcommand{\CanonicalMbHadInf}{\result{\ensuremath{1.552^{+0.010}_{-0.010}}}}
\newcommand{\MmaxHadInf}{\result{\ensuremath{2.233^{+0.098}_{-0.098}}}}

\newcommand{\CanonicalRhocHadInf}{\result{\ensuremath{6.87^{+0.85}_{-0.84}\times10^{14}}}}
\newcommand{\POneRhonucHadInf}{\result{\ensuremath{ 6.07^{+1.29}_{-1.29}\times10^{33}}}}
\newcommand{\PTwoRhonucHadInf}{\result{\ensuremath{ 4.71^{+0.96}_{-0.84}\times10^{34}}}}
\newcommand{\PSixRhonucHadInf}{\result{\ensuremath{11.81^{+1.45}_{-1.46}\times10^{35}}}}

\newcommand{\CanonicalLHypInf}{\result{\ensuremath{650^{+113}_{-134}}}}
\newcommand{\CanonicalRHypInf}{\result{\ensuremath{13.23^{+0.37}_{-0.43}}}}
\newcommand{\CanonicalIHypInf}{\result{\ensuremath{1.67^{+0.08}_{-0.10}}}}
\newcommand{\CanonicalMbHypInf}{\result{\ensuremath{1.589^{+0.008}_{-0.007}}}}
\newcommand{\MmaxHypInf}{\result{\ensuremath{2.017^{+0.071}_{-0.084}}}}

\newcommand{\CanonicalRhocHypInf}{\result{\ensuremath{6.71^{+0.72}_{-0.64}\times10^{14}}}}
\newcommand{\POneRhonucHypInf}{\result{\ensuremath{5.86^{+1.03}_{-1.01}\times10^{33}}}}
\newcommand{\PTwoRhonucHypInf}{\result{\ensuremath{4.84^{+0.59}_{-0.66}\times10^{34}}}}
\newcommand{\PSixRhonucHypInf}{\result{\ensuremath{6.70^{+0.64}_{-0.60}\times10^{35}}}}

\newcommand{\CanonicalLQrkInf}{\result{\ensuremath{427^{+125}_{-143}}}}
\newcommand{\CanonicalRQrkInf}{\result{\ensuremath{12.14^{+0.59}_{-0.72}}}}
\newcommand{\CanonicalIQrkInf}{\result{\ensuremath{1.50^{+0.11}_{-0.14}}}}
\newcommand{\CanonicalMbQrkInf}{\result{\ensuremath{1.568^{+0.017}_{-0.016}}}}
\newcommand{\MmaxQrkInf}{\result{\ensuremath{1.978^{+0.082}_{-0.048}}}}

\newcommand{\CanonicalRhocQrkInf}{\result{\ensuremath{7.47^{+1.15}_{-1.02}\times10^{14}}}}
\newcommand{\POneRhonucQrkInf}{\result{\ensuremath{3.18^{+1.35}_{-1.24\times10^{33}}}}}
\newcommand{\PTwoRhonucQrkInf}{\result{\ensuremath{4.06^{+1.14}_{-1.28}\times10^{34}}}}
\newcommand{\PSixRhonucQrkInf}{\result{\ensuremath{6.55^{+0.87}_{-0.89}\times10^{35}}}}

\newcommand{\MOneHadInf}{\result{\ensuremath{1.45^{+0.10}_{-0.09}}}}
\newcommand{\MTwoHadInf}{\result{\ensuremath{1.28^{+0.08}_{-0.08}}}}
\newcommand{\LOneHadInf}{\result{\ensuremath{444^{+222}_{-250}}}}
\newcommand{\LTwoHadInf}{\result{\ensuremath{961^{+532}_{-382}}}}
\newcommand{\LTildeHadInf}{\result{\ensuremath{658^{+188}_{-156}}}}
\newcommand{\ROneHadInf}{\result{\ensuremath{12.96^{+0.57}_{-0.55}}}}
\newcommand{\RTwoHadInf}{\result{\ensuremath{13.03^{+0.55}_{-0.57}}}}
\newcommand{\RhocOneHadInf}{\result{\ensuremath{7.07^{+1.10}_{-0.99}}}} % \times10^{14}
\newcommand{\RhocTwoHadInf}{\result{\ensuremath{6.41^{+0.77}_{-0.85}}}} % \times10^{14}

\newcommand{\MOneHadAgn}{\result{\ensuremath{1.50^{+0.13}_{-0.13}}}}
\newcommand{\MTwoHadAgn}{\result{\ensuremath{1.25^{+0.11}_{-0.10}}}}
\newcommand{\LOneHadAgn}{\result{\ensuremath{124^{+181}_{-103}}}}
\newcommand{\LTwoHadAgn}{\result{\ensuremath{379^{+402}_{-257}}}}
\newcommand{\LTildeHadAgn}{\result{\ensuremath{216^{+266}_{-141}}}}
\newcommand{\ROneHadAgn}{\result{\ensuremath{10.66^{+1.66}_{-1.35}}}}
\newcommand{\RTwoHadAgn}{\result{\ensuremath{10.59^{+1.79}_{-1.57}}}}
\newcommand{\RhocOneHadAgn}{\result{\ensuremath{10.38^{+3.35}_{-3.76}}}} % \times10^{14}
\newcommand{\RhocTwoHadAgn}{\result{\ensuremath{ 9.36^{+2.59}_{-3.21}}}} % \times10^{14}

\newcommand{\MOneHypInf}{\result{\ensuremath{1.46^{+0.08}_{-0.09}}}}
\newcommand{\MTwoHypInf}{\result{\ensuremath{1.28^{+0.08}_{-0.07}}}}
\newcommand{\LOneHypInf}{\result{\ensuremath{502^{+294}_{-187}}}}
\newcommand{\LTwoHypInf}{\result{\ensuremath{1114^{+495}_{-414}}}}
\newcommand{\LTildeHypInf}{\result{\ensuremath{760^{+131}_{-150}}}}
\newcommand{\ROneHypInf}{\result{\ensuremath{13.20^{+0.39}_{-0.43}}}}
\newcommand{\RTwoHypInf}{\result{\ensuremath{13.28^{+0.34}_{-0.45}}}}
\newcommand{\RhocOneHypInf}{\result{\ensuremath{7.02^{+0.87}_{-0.91}}}} % \times10^{14}
\newcommand{\RhocTwoHypInf}{\result{\ensuremath{6.19^{+0.71}_{-0.65}}}} % \times10^{14}

\newcommand{\MOneHypAgn}{\result{\ensuremath{1.47^{+0.12}_{-0.11}}}}
\newcommand{\MTwoHypAgn}{\result{\ensuremath{1.27^{+0.10}_{-0.09}}}}
\newcommand{\LOneHypAgn}{\result{\ensuremath{268^{+324}_{-229}}}}
\newcommand{\LTwoHypAgn}{\result{\ensuremath{690^{+600}_{-522}}}}
\newcommand{\LTildeHypAgn}{\result{\ensuremath{466^{+337}_{-337}}}}
\newcommand{\ROneHypAgn}{\result{\ensuremath{12.14^{+1.39}_{-1.90}}}}
\newcommand{\RTwoHypAgn}{\result{\ensuremath{12.15^{+1.44}_{-2.07}}}}
\newcommand{\RhocOneHypAgn}{\result{\ensuremath{8.29^{+3.51}_{-2.48}}}} % \times10^{14}
\newcommand{\RhocTwoHypAgn}{\result{\ensuremath{7.32^{+3.11}_{-1.96}}}} % \times10^{14}

\newcommand{\MOneQrkInf}{\result{\ensuremath{1.46^{+0.12}_{-0.10}}}}
\newcommand{\MTwoQrkInf}{\result{\ensuremath{1.27^{+0.08}_{-0.10}}}}
\newcommand{\LOneQrkInf}{\result{\ensuremath{315^{+210}_{-179}}}}
\newcommand{\LTwoQrkInf}{\result{\ensuremath{742^{+425}_{-328}}}}
\newcommand{\LTildeQrkInf}{\result{\ensuremath{496^{+145}_{-166}}}}
\newcommand{\ROneQrkInf}{\result{\ensuremath{12.14^{+0.63}_{-0.68}}}}
\newcommand{\RTwoQrkInf}{\result{\ensuremath{12.10^{+0.59}_{-0.76}}}}
\newcommand{\RhocOneQrkInf}{\result{\ensuremath{7.83^{+1.44}_{-1.30}}}} % \times10^{14}
\newcommand{\RhocTwoQrkInf}{\result{\ensuremath{6.94^{+1.06}_{-0.94}}}} % \times10^{14}

\newcommand{\MOneQrkAgn}{\result{\ensuremath{1.49^{+0.13}_{-0.12}}}}
\newcommand{\MTwoQrkAgn}{\result{\ensuremath{1.25^{+0.10}_{-0.10}}}}
\newcommand{\LOneQrkAgn}{\result{\ensuremath{162^{+260}_{-135}}}}
\newcommand{\LTwoQrkAgn}{\result{\ensuremath{456^{+483}_{-310}}}}
\newcommand{\LTildeQrkAgn}{\result{\ensuremath{262^{+335}_{-162}}}}
\newcommand{\ROneQrkAgn}{\result{\ensuremath{10.94^{+1.80}_{-1.27}}}}
\newcommand{\RTwoQrkAgn}{\result{\ensuremath{10.84^{+1.88}_{-1.45}}}}
\newcommand{\RhocOneQrkAgn}{\result{\ensuremath{9.46^{+3.02}_{-3.52}}}} % \times10^{14}
\newcommand{\RhocTwoQrkAgn}{\result{\ensuremath{8.55^{+2.41}_{-2.90}}}} % \times10^{14}

\newcommand{\apjl}{Astrophys.~J.~Lett.}
\newcommand{\procspie}{Proc.~SPIE}
\newcommand{\aap}{Astron.~Astrophys.}
\newcommand{\mnras}{Mon.~Not.~R.~Astron.~Soc.}
\newcommand{\nphysa}{Nucl.~Phys.~A}
\newcommand{\apjs}{Astrophys.~J.~Suppl.}

\begin{document}
%-------------------------------------------------

\title{
Nonparametric inference of neutron star composition, equation of state, \\ and maximum mass with GW170817
}

\author{Reed Essick}
\email{reedessick@kicp.uchicago.edu}
\affiliation{Kavli Institute for Cosmological Physics, The University of Chicago, 5640 South Ellis Avenue, Chicago, Illinois, 60637, USA}

\author{Philippe Landry}
\email{landryp@uchicago.edu}
\affiliation{Enrico Fermi Institute and Kavli Institute for Cosmological Physics, The University of Chicago, 5640 South Ellis Avenue, Chicago, Illinois, 60637, USA}

\author{Daniel E. Holz}
\email{holz@uchicago.edu}
\affiliation{Enrico Fermi Institute, Department of Physics, Department of Astronomy and Astrophysics,\\and Kavli Institute for Cosmological Physics, The University of Chicago, Chicago, IL 60637, USA}

\date{\today}

\begin{abstract}
The detection of GW170817 in gravitational waves provides unprecedented constraints on the equation of state (\EOS) of the ultra-dense matter within the cores of neutron stars (\NSs).
We extend the nonparametric analysis first introduced in Landry \& Essick (2019), and confirm that GW170817 favors soft \EOSs.
We infer macroscopic observables for a canonical $1.4\,M_\odot$ \NS, including the tidal deformability $\Lambda_{1.4}=\CanonicalLAgnostic$ ($\CanonicalLInformed$) and radius $R_{1.4}=\CanonicalRAgnostic$ ($\CanonicalRInformed$) km, as well as the maximum mass for nonrotating \NSs, $\Mmax=\MmaxAgnostic$ ($\MmaxInformed$)$\,M_\odot$, with nonparametric priors loosely (tightly) constrained to resemble candidate \EOSs~from the literature.
Furthermore, we find weak evidence that GW170817 involved at least one \NS~based on gravitational-wave data alone ($B^{\mathrm{\NS}}_{\BBH}=\BayesNSBHHispinAgnostic$), consistent with the observation of electromagnetic counterparts.
We also investigate GW170817's implications for the maximum spin frequency of millisecond pulsars, and find that the fastest known pulsar is spinning at more than \result{50\%} of its breakup frequency at 90\% confidence.
We additionally find modest evidence in favor of quark matter within \NSs, and GW170817 favors the presence of at least one disconnected hybrid star branch in the mass--radius relation over a single stable branch by a factor of $\BayesBranchesApprox$.
Assuming there are multiple stable branches, we find a suggestive posterior preference for a sharp softening around nuclear density followed by stiffening around twice nuclear density, consistent with a strong first-order phase transition.
While the statistical evidence in favor of new physics within \NS~cores remains tenuous with GW170817 alone, these tantalizing hints reemphasize the promise of gravitational waves for constraining the supranuclear \EOS.
\end{abstract}

\maketitle

%-------------------------------------------------

\section{Introduction}
\label{sec:intro}

The observation of gravitational waves (\GWs) from GW170817 \cite{GW170817discovery}, a coalescing compact binary with an electromagnetic counterpart, has greatly advanced the study of nuclear matter at extreme densities.
Changes in the orbital phasing due to the components' mutual tidal interaction leave a detectable imprint in the \GW~signal \cite{Hinderer2010,ReadBaiotti2013,DelPozzo2013,WadeCreighton2014,DamourNagar2012}, and several studies have exploited this fact to infer the tidal deformabilities of the compact objects involved in the coalescence \cite{GW170817properties, De2018, GW170817eos, Landry2019}.
Additionally, the observations of counterparts across the electromagnetic spectrum provide complementary constraints on tidal effects \cite{Margalit2017, Radice2018a, Radice2018b, Coughlin2018}.
As the advanced LIGO~\cite{LIGO} and Virgo~\cite{Virgo} detectors continue to operate, further observations of \NS~coalescences (e.g. Refs.~\cite{S190425zGCN, S190814bvGCN}) will add to our knowledge of the supranuclear equation of state (\EOS) \cite{LackeyWade2015,McNeilForbes2019}.

Landry \& Essick \cite{Landry2019} (hereafter \LandryEssick) recently introduced a nonparametric method for inferring the \NS~\EOS~from \GW~observations based on Gaussian processes that automatically incorporate physical constraints, like causality and thermodynamic stability.
We extend their methodology, apply it to GW170817, and obtain updated constraints on the \EOS, as well as new bounds on several derived quantities.
This includes testing hypotheses about the composition of dense matter and support for hybrid stars for the first time, as well as revisiting the consistency of observed pulsar spins with their rotational breakup frequencies.

\LandryEssick~discussed the advantages of nonparametric analyses over parametric inference schemes with a finite number of parameters.
Our updated analysis continues to avoid the kind of modeling systematics inherent to a coarse parametric representation of the \EOS's unknown functional form, and we additionally improve the construction of the nonparametric \EOS~prior in several ways, reducing the impact of \textit{ad-hoc} choices made within \LandryEssick.
Whereas \LandryEssick~chose hyperparameters for their Gaussian processes by hand when constructing their priors, we select them by finding hyperparameter sets that optimally reproduce the variability seen in a training set of candidate \EOSs.
We further sample over a mixture model of such sets, representing the overall process as a weighted sum over many individual Gaussian processes.
Our priors, like \LandryEssick's, naturally incorporate variable uncertainty in the \EOS~at different pressures, including tight constraints at low densities, where nuclear matter is better understood, and broader uncertainties at high densities.
We also train our Gaussian processes on 50 tabulated candidate \EOSs~(as opposed to the 7 used in \LandryEssick) and subdivide the resulting priors according to the composition of the \EOSs~on which they were trained.
This elucidates finer-grained questions about \NS~composition while incorporating a broader range of theoretical expectations.
Moreover, like Ref.~\cite{Gamba2019}, our analysis marginalizes over several possible crust \EOSs~to account for the (relatively small) uncertainty in the \NS~\EOS~at low densities.
Leveraging this, we present new results based on the inferred posterior process for the \EOS, including both posterior distributions for macroscopic observables associated with GW170817 itself and functional relations between generic \NS~observables.

Consistent with previous studies~\cite{GW170817eos, GW170817modelselection, Landry2019, De2018}, we find that GW170817 favors relatively soft \EOSs, assuming the system's components were both slowly spinning \NSs~\cite{GW170817properties}.
This manifests as an overall posterior preference for lower pressures at and above nuclear density ($\rhonuc=2.8\times10^{14}\,\mathrm{g}/\mathrm{cm}^3$), as well as smaller radii ($R$), tidal deformabilities ($\Lambda$), and maximum masses for nonrotating \NSs~(\Mmax).
Our results are conditioned on the existence of $\approx 2M_\odot$ pulsars~\cite{Antoniadis2013, Cromartie2019} \textit{a priori}, as we retain only \EOSs~drawn from our prior that support \NSs~of at least $1.93\,M_\odot$.
All our conclusions, then, depend on \GW~and pulsar data, and our Bayes factors compare our \GPs~conditioned on both to those~conditioned only on pulsar observations.
Like \LandryEssick, we derive results using both \emph{model-agnostic} and \emph{model-informed} priors, reflecting different amounts of relative \textit{a priori} confidence in candidate \EOSs~from the literature.
Our \emph{informed} prior is conditioned to closely emulate the behavior of \EOSs~proposed in the literature, whereas our \emph{agnostic} prior generates much more diverse \EOS~behavior and is not tightly constrained by the \EOS~upon which it was conditioned.
With the \emph{agnostic} (\emph{informed}) prior, we infer median and 90\% highest-probability-density credible regions for the macroscopic observables of a canonical 1.4$\,M_\odot$ \NS: $\Lambda_{1.4}=\CanonicalLAgnostic$ ($\CanonicalLInformed$) and $R_{1.4}=\CanonicalRAgnostic$ ($\CanonicalRInformed$) km, with $\Mmax=\MmaxAgnostic$ ($\MmaxInformed$)$\,M_\odot$, after marginalizing over \EOS~composition.
These results are broadly consistent with \LandryEssick~and other studies.

By constructing separate priors for different \EOS~compositions, our updated analysis shows that GW170817 weakly favors \EOSs~that contain quark matter: $P(\mathrm{Quark}|\mathrm{data}) = \PQrkInformed$ assuming equal prior odds for hadronic, hyperonic, and quark compositions with our \emph{informed} prior.
Remarkably, with the \emph{agnostic} prior, we find that GW170817 modestly favors \EOSs~that support a disconnected hybrid star branch, one signature of a strong first-order phase transition.
Among such \EOSs, GW170817 suggests a possible phase transition with onset density between \rhonuc~and 2\rhonuc, in agreement with chiral effective field theory predictions for the breakdown of perturbations off of asymmetric nuclear matter~\cite{Tews2018a, Tews2018b, Tews2019, Furnstahl2015}.
Our nonparametric inference attaches no \textit{a priori} significance to these particular densities;
the preference observed \textit{a posteriori} is entirely driven by data from GW170817 and \Mmax~constraints from observations of massive pulsars.
While these results are far from conclusive, they demonstrate the extent of information available from GW observations and hint at new physics within \NS~cores.

We also reexamine limits on \NS~spin based on the \EOS, finding maximum dimensionless spins \result{$\chi_\mathrm{max}\lesssim0.5$} for masses $M \gtrsim M_{\odot}$.
We find that the fastest known pulsar, J1748-2446ad~\cite{Hessels2006}, which does not have a precisely measured mass, rotates at \result{$\gtrsim1/2$} its breakup frequency for the same mass range at 90\% confidence.

We additionally compute the relative marginal likelihoods for different progenitor systems, e.g.~binary \NS~(\BNS) vs.~NS-black hole (\NSBH), finding a preference for progenitor systems containing at least one \NS~compared to a binary \BH~(\BBH) by a factor of $\BayesNSBHHispinAgnostic$ while making minimal assumptions about the components' spins, in agreement with the observation of electromagnetic counterparts~\cite{GW170817mma, GW170817grb}.
Interestingly, we find a further slight preference for the lighter component to be a \BH, with $B^\NSBH_\BNS = \BayesNSBHBNSAgnostic$ for our \emph{agnostic} prior.
This is likely due to GW170817 favoring relatively small $\Lambda_2$, which is slightly more consistent with a \BH~($\Lambda_2=0$) than the large $\Lambda_2$ required by most \NS~\EOSs.

As with \LandryEssick, our \emph{agnostic} and \emph{informed} results bracket other results in the literature.
For example, Ref.~\cite{GW170817eos} constrained the radii of each of GW170817's components to be \externalresult{$11.9^{+1.4}_{-1.4}$} km, while Ref.~\cite{De2018} assumed $R_1=R_2$ and found a radius of \externalresult{8.9--13.2} km.
Similarly, Ref.~\cite{GW170817properties} constrained \externalresult{$\tilde{\Lambda}=300^{+420}_{-230}$} and Ref.~\cite{De2018} found \externalresult{$\tilde{\Lambda} =  222^{+420}_{-138}$} assuming a uniform component mass prior.
We infer $\tilde{\Lambda}=\LTildeAgnostic$ (\LTildeInformed) with the \emph{agnostic} (\emph{informed}) prior.
Although no previous constraints on \NS~composition are available, we note that Ref.~\cite{GW170817modelselection} calculated Bayes factors between \BNS~and \BBH~models for individual candidate \EOSs, with \externalresult{$\ln B^{\BNS}_{\BBH} \lesssim 2$} using a broad mass prior and the majority of candidate \EOSs~yielding \externalresult{$\ln B^\BNS_\BBH \sim 0$}.
This is in good agreement with our \emph{agnostic} estimate of $\ln B^{\BNS}_{\BBH}(\chi_i\leq0.89)=\lnBayesBNSBBHHispinAgnostic$.
Where comparable, our findings are generally in good agreement with existing results in the literature.
We additionally present a number of novel results, including the first evidence for \NS~composition and the existence of hybrid-star branches, and revisit spin constraints for rapidly rotating pulsars in light of GW170817.

We review the nonparametric inference introduced in \LandryEssick~in Section~\ref{sec:inference}, including descriptions of our improvements.
Section~\ref{sec:setup} describes the priors constructed for this work.
Using publicly available posterior samples~\cite{GW170817samples} from a study of GW170817's source properties \cite{GW170817properties}, Section~\ref{sec:results} presents \textit{a posteriori} constraints obtained for macroscopic observables associated with GW170817, such as the component masses and tidal deformabilities, while Section~\ref{sec:ilove} presents constraints for relationships between macroscopic observables, applicable to systems besides GW170817.
Section~\ref{sec:composition model selection} describes our inference over \NS~compositions, and we conclude in Section~\ref{sec:disc}.

%-------------------------------------------------
\section{Nonparametric Inference of the Equation of State}
\label{sec:inference}

We extend the nonparametric inference based on Gaussian processes (\GPs) detailed in \LandryEssick, and refer readers to that paper for a pedagogical introduction to \GPs, their use in our analysis, and associated notation.
Nonetheless, we provide a brief overview in what follows.

A \GP~assumes Gaussian correlations between functional degrees of freedom, described by a mean and covariance.
By conditioning a joint process on the observed data and assuming a functional form for the covariance, we obtain a process for infinitely many degrees of freedom based on a finite set of known data, and the complexity of the resulting nonparametric model can naturally scale with the amount of available data.
A thorough description is available in~\cite{Rasmussen2006}, but the key insight is that the probability distribution of a functional degree of freedom ($f$) given a corresponding abscissa ($x$) and known data ($f_\ast$, $x_\ast$) is
\begin{equation}\label{eqn:conditioning}
    P(f|f_\ast, x, x_\ast; \vec{\sigma}) = \frac{P(f, f_\ast|x, x_\ast; \vec{\sigma})}{P(f_\ast|x_\ast; \vec{\sigma})}
\end{equation}
assuming
\begin{equation}
    P(f, f_\ast|x, x_\ast, \vec{\sigma}) = \mathcal{N}\big(\mu(x_i), K(x_i, x_j; \vec{\sigma})\big) ,
\end{equation}
where $\mu(x_i)$ is the mean and $K(x_i, x_j; \vec{\sigma})$ the covariance of a multivariate normal distribution.
$K$ is a function of the \emph{hyperparameters} $\vec{\sigma}$.
We use the squared-exponential kernel
\begin{equation}
    K_\mathrm{se}(x_i, x_j; \sigma, l) = \sigma^2 \exp\left(-\frac{(x_i-x_j)^2}{2l^2}\right),
\end{equation}
which models correlations between neighboring functional degrees of freedom, the white-noise kernel
\begin{equation}
    K_\mathrm{wn}(x_i, x_j; \sigma_\mathrm{obs}) = \sigma_\mathrm{obs}(x_i) \delta(x_i - x_j),
\end{equation}
which models uncertainty at each point, and a scaled covariance between input models
\begin{multline}
    K_\mathrm{mv}(x_i, x_j; m) = \\
        m^2 \left(\frac{1}{N_A}\sum\limits_{a \in A} C^{(a)}_{ij} \quad + \right. \quad \quad \quad \quad \quad \quad \quad \quad \quad \quad \quad \quad \quad \quad \\
        \left. \frac{1}{N_A}\sum\limits_{a \in A} \left(\mu^{(a)}_i - \bar{\mu}^{(A)}_i\right)\left(\mu^{(a)}_j - \bar{\mu}^{(A)}_j\right)\right)
\end{multline}
where $A$ is the set of $N_A$ input models,
$\mu^{(a)}$ and $C^{(a)}$ are the mean and covariance of the process for model $a$,
\begin{equation}
    \bar{\mu}^{(A)}_i = \frac{1}{N_A}\sum\limits_{a \in A} \mu^{(a)}_i,
\end{equation}
and $m$ scales the relative importance of this model covariance.
$K_\mathrm{mv}$, like $K_\mathrm{wn}$, represents theoretical uncertainty.
We note that \LandryEssick~used a simplified version of $K_\mathrm{mv}$ which only included the diagonal components of the covariance matrix.

We generate \GPs~for an auxiliary variable
\begin{equation}
    \phi = \log\left( c^2 \frac{d\varepsilon}{dp} - 1\right)
\end{equation}
conditioned on tabulated \EOSs~from the literature, where $\varepsilon$ is the total energy density and $p$ the pressure.
Any realization of $\phi$ will automatically satisfy both causality and thermodynamic-stability constraints ($0 \leq c_s^2 = dp/d\varepsilon \leq c^2$).
Our method employs \GPs~for two main purposes: mapping irregularly sampled tabulated data for $\varepsilon(p)$ into a regularly sampled process for $\phi(p)$ while self-consistently computing the uncertainty in that mapping, and emulating the behavior seen in tabulated \EOSs~to generate synthetic \EOSs~which resemble models from the literature.
This work differs from \LandryEssick~in that we construct mixture models of \GPs~instead of relying on a single set of hyperparameters.
That is,
\begin{equation}
    \phi \sim \sum\limits_i w_i P(\phi|\vec{\sigma}_i)
\end{equation}
where $\vec{\sigma}_i$ and $w_i$ are the hyperparameters and weight associated with the $i^\mathrm{th}$ element of the mixture model.

We constrain our processes to approximately match known low-pressure physics with an additional white-noise variant.
This forces all realizations of the conditioned process to approach a constant value (chosen to be $\phi\rightarrow\phi_0=6$ based on $\texttt{sly}$~\cite{Douchin2001}) with a white-noise scaling parameter
\begin{equation}
    \sigma_\mathrm{obs}(p) = \left(\frac{p}{p_\mathrm{ref}}\right)^n .
\end{equation}
At low pressures, $\sigma_\mathrm{obs}\rightarrow0$, forcing the conditioned process to approach $\phi_0$ while imposing no constraint when $p \gg p_\mathrm{ref}$.
Here we set $p_\mathrm{ref} = 5.4\times10^{31}\, \mathrm{dyn}/\mathrm{cm}^{2}$ and $n=5$, \textit{ad hoc} choices with negligible impact on the resulting \EOS, as we match the GP realizations onto a fixed model for the low-density crust well above $p_{\rm ref}$.
Improved matching conditions that incorporate the expected uncertainty from first-principles theoretical calculations, such as those in Refs.~\cite{Furnstahl2015, Hebeler2013, Tews2013, Tews2018a}, may further enhance our analysis.
Similarly, matching to known high-density behavior, like $c_s^2 \rightarrow c^2/3$ for hyperrelativistic matter~\cite{Fraga2016, Kurkela2014}, could prove interesting.
However, we leave this to future work.

In addition to these technical improvements, we extend \LandryEssick's analysis by conditioning our GPs on more tabulated \EOSs~(50 instead of 7) with a broader range of phenomenology and compositions.
We also subdivide our \emph{agnostic} and \emph{informed} priors according to the composition of the input candidate \EOSs~(hadronic $npe\mu$ matter, hyperonic $npe\mu{Y}$ matter, or $npe\mu(Y)Q$ quark matter) allowing for model selection between different \NS~constituents.

%------------------------
\subsection{Constructing nonparametric priors}
\label{sec:setup}

We construct several priors using the candidate \EOSs~listed in Table~\ref{table:tabulated eos}.
In order to fairly weight the importance of each input \EOS, we group them by composition and underlying family of nuclear effective forces, generating representative processes for each family separately and then weighting the resulting \GPs~equally.
This is done hierarchically in order to synthesize overarching \textit{agnostic} and \textit{informed} priors, as well as priors that demonstrate behavior characteristic of \EOSs~with a particular composition.

We map $\varepsilon$ to $\phi$ for each \EOS, modeling each one's covariance matrix separately with a squared-exponential kernel with a small white-noise term ($\sigma \gg \sigma_\mathrm{obs}$), and then generate a sequence of overarching \GPs~which emulate the behavior observed between different \EOSs~using a combination of squared-exponential, white-noise, and model covariance kernels.
As in \LandryEssick, we take the mean of the joint process before conditioning to be a low-order polynomial fit to the input data.

We optimize the hyperparameters used to generate our priors with a cross-validation likelihood
\begin{equation}\label{eqn:cv likelihood}
    P_\mathrm{CV}(\{\varepsilon\}_{A}|p,\vec{\sigma}) = \prod\limits_{a \in A} P(\varepsilon^{(a)}|\{\varepsilon\}_{A \setminus a},p,\vec{\sigma}) ,
\end{equation}
where $\{\varepsilon\}_A$ is the set of all \EOSs~and $\{\varepsilon\}_{A \setminus a}$ is the set of all \EOSs~in $A$ except $\varepsilon^{(a)}$, and obtain processes that emulate the behavior seen within each composition separately.
However, instead of selecting a single set of optimal hyperparameters for each composition, we instead create mixture models by drawing many sets of hyperparameters from $P_\mathrm{CV}$ so that
\begin{multline}\label{eqn:gp mixture model}
    P(\varepsilon|\{\varepsilon\}_A, p) \\
    \propto \int d\vec{\sigma}\, P(\vec{\sigma}) P_\mathrm{CV}(\{\varepsilon\}_A|\vec{\sigma})^\beta P(\varepsilon|\{\varepsilon\}_A,p,\vec{\sigma}) \\
    \approx \left. \frac{1}{N}\sum\limits_i P_\mathrm{CV}(\{\varepsilon\}_A|\vec{\sigma}_i)^\beta P(\varepsilon|\{\varepsilon\}_A,p,\vec{\sigma}_i)\quad \right| \quad \vec{\sigma}_i \sim P(\vec{\sigma})
\end{multline}
where $\beta=1/T$ is an inverse temperature.

\begin{figure*}
    \begin{minipage}{0.49\textwidth}
        \emph{model-agnostic} \\
        \includegraphics[width=1.0\textwidth]{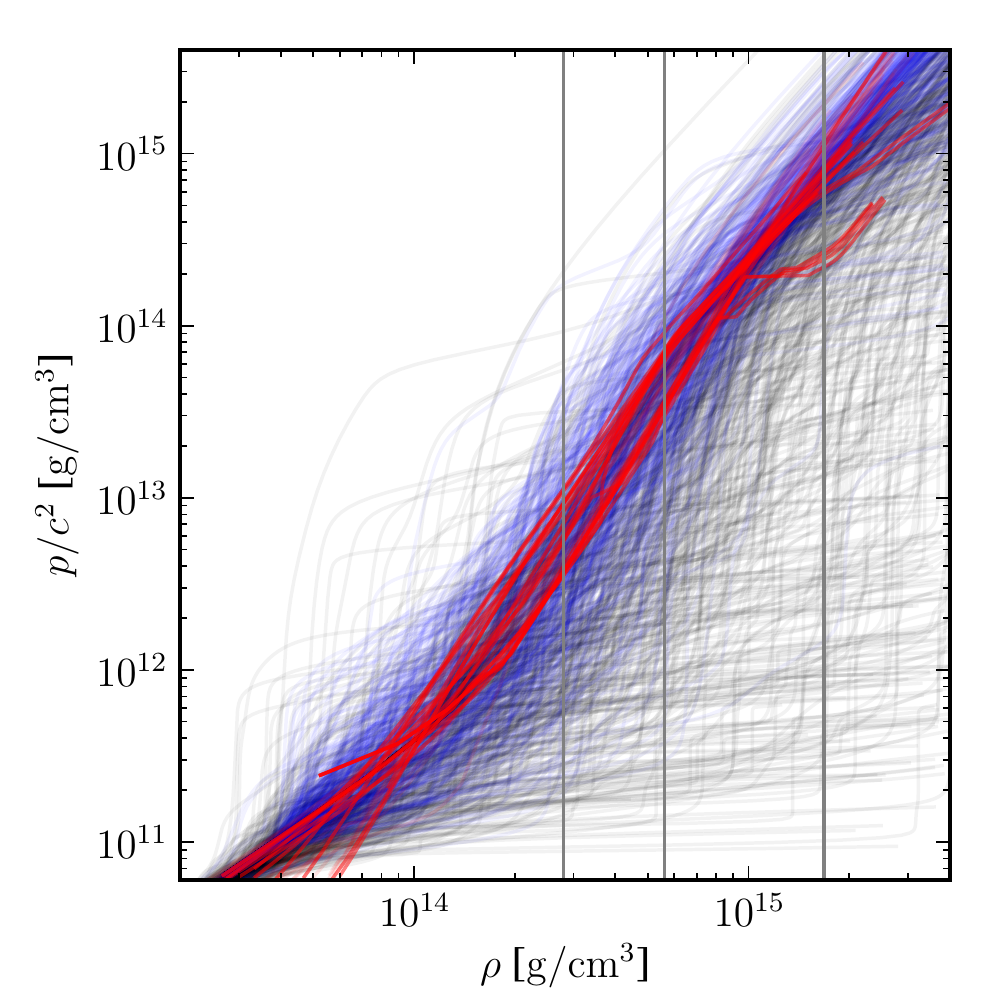}
    \end{minipage}
    \begin{minipage}{0.49\textwidth}
        \emph{model-informed} \\
        \includegraphics[width=1.0\textwidth]{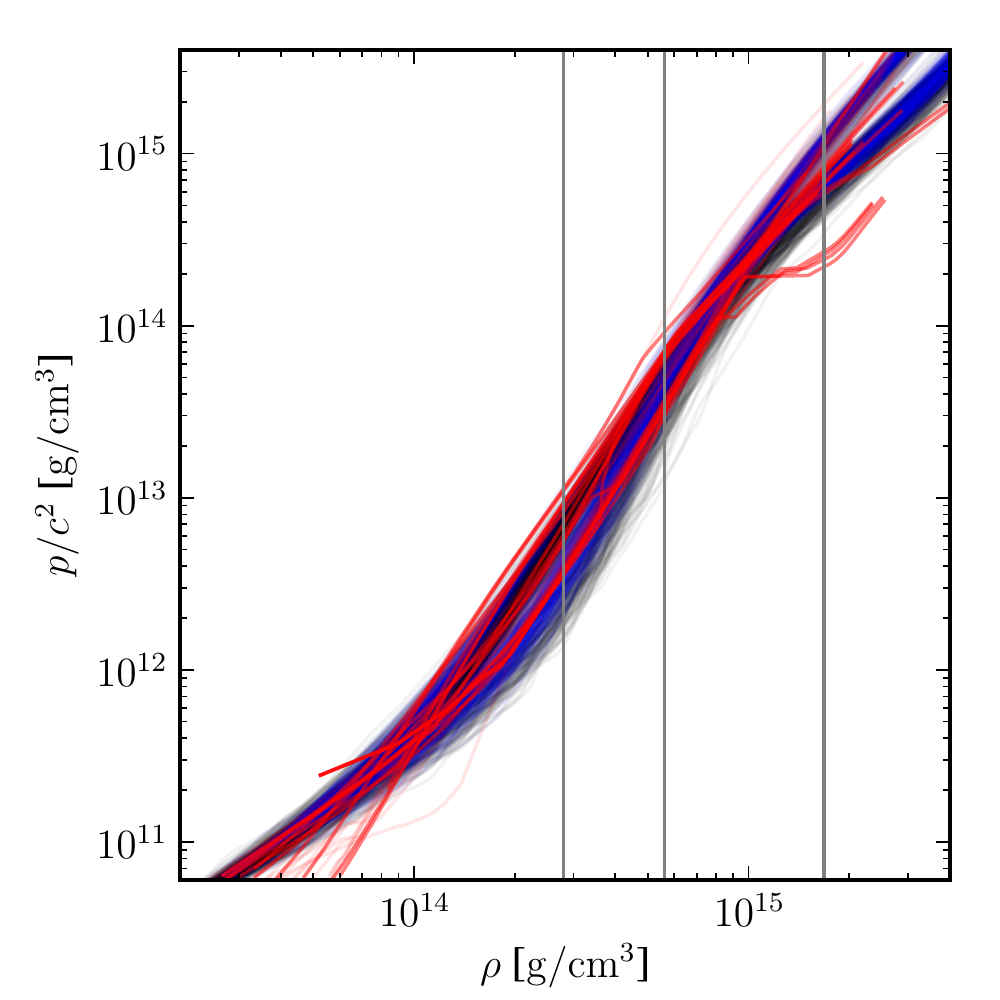}
    \end{minipage}
    \caption{
        Example synthetic \EOSs~drawn from our (\emph{left}) \emph{agnostic} and (\emph{right}) \emph{informed} nonparametric priors, constructed as mixture models with equal prior odds for hadronic, hyperonic, and quark compositions.
        Draws from the prior are colored according to the maximum nonrotating \NS~mass they support: \emph{blue} for \Mmax $\,\geq$ 1.93$\,M_\odot$, and \emph{black} otherwise.
        Candidate \EOSs~from the literature, used as input for our \GPs, are shown in \emph{red} (see Table~\ref{table:tabulated eos}). Vertical lines indicate once, twice and six times nuclear saturation density.
    }
    \label{fig:priors marg}
\end{figure*}

In the limit $T\rightarrow1$, we weight each set of hyperparameters by the cross-validation likelihood.
We call the resulting \GP-mixure model the \emph{model-informed} prior.
In the limit $T\rightarrow\infty$, we weight each set of hyperparameters equally, subject to the hyperprior $P(\vec{\sigma})$.
This produces a process much less constrained by the input \EOSs, which we refer to as the \emph{model-agnostic} prior.
We sample logarithmically in $\sigma$, $\sigma_\mathrm{obs}$, and $m$ while sampling linearly in $l$.
The precise choice of $P(\vec{\sigma})$ does not strongly affect the \emph{model-informed} prior, but can modify the behavior of the \emph{model-agnostic} prior.
It is also worth noting that $\beta$ provides a natural way to tune the \textit{a priori} degree of belief placed on published theoretical models, and our choices for the \emph{informed} and \emph{agnostic} priors are not the only ones possible.
Example synthetic \EOSs~are shown Figure~\ref{fig:priors marg}.

%------------------------
\subsection{Sampling from nonparametric priors}
\label{sec:integration}

We obtain a large number of synthetic \EOSs~by sampling from each process for $\phi$.
We integrate each realization $\phi^{(\alpha)}(p)$ to obtain the associated synthetic \EOS~$\varepsilon^{(\alpha)}(p)$.
This includes setting the initial value for the integration, which is done by matching to a crust \EOS~at a particular pressure.
To marginalize over the (relatively small) theoretical uncertainty in the microscopic description of the crust, we randomly match each $\varepsilon^{(\alpha)}$ to the low-density model used in one of either \texttt{sly}~\cite{Douchin2001}, \texttt{eng}~\cite{Engvik1995}, or \texttt{hqc18}~\cite{Baym2019}.
Furthermore, the precise pressure at which we match the crust \EOS~is allowed to vary by approximately an order of magnitude.
This procedure generates \EOSs~which span all densities relevant for \NS~structure.

We integrate each synthetic \EOS~to obtain the corresponding relationships between macroscopic observables, such as the mass, radius, and tidal deformability.
As in \LandryEssick, we additionally require that each $\varepsilon^{(\alpha)}$ support at least a $1.93\,M_\odot$ star based on the existence of massive pulsars (J0348+0432~\cite{Antoniadis2013}), discarding any synthetic \EOS~that does not.
We also discard any \EOS~that supports a spurious branch of \NSs~with $M\gtrsim M_\odot$ at subnuclear densities $\rho_c < 0.8\rhonuc$, so as to exclude particularly large excursions in pressure just above the crust-core interface.
We obtain marginal likelihoods, posterior distributions, and posterior processes by Monte-Carlo sampling from the prior, weighting each sample with a Gaussian kernel density estimate (\KDE) for the \GW~likelihood generated with publicly available samples~\cite{GW170817samples}:
\begin{widetext}
\begin{align}
    P(d|\varepsilon^{(\alpha)}, \mathcal{H}) & = \int dM_1 dM_2\, p(M_1, M_2|\mathcal{H})\, \mathcal{L}\left(\mathrm{data} \left| M_1, M_2, \Lambda^{(\alpha)}(M_1), \Lambda^{(\alpha)}(M_2)\right.\right) \\
                                             & \approx \frac{1}{N_i} \sum\limits_i^{N_i} \mathcal{L}\left(\mathrm{data} \left| M_1^{(i)}, M_2^{(i)}, \Lambda^{(\alpha)}(M_1^{(i)}), \Lambda^{(\alpha)}(M_2^{(i)}) \right.\right) \quad \left| \quad M_1^{(i)}, M_2^{(i)} \sim P(M_1, M_2|\mathcal{H}) \right. ,
\end{align}
where $\Lambda^{(\alpha)}$ is the mass-tidal deformability relation implicitly defined by $\varepsilon^{(\alpha)}$.
It is worth noting that several sets of samples are publicly accessible.
Our specific choice is not expected to significantly affect our conclusions, although our precise quantitative results will depend on issues like waveform systematics discussed in Ref.~\cite{GWTC1}.
Drawing $\varepsilon^{(\alpha)}$ from our prior and associating this marginal likelihood with each sample generates the posterior process.
This also allows us to immediately estimate the evidence for each prior, up to a common normalization constant:
\begin{equation}
    P(d|\{\varepsilon\}_A, \mathcal{H}) \approx \frac{1}{N_\alpha}\sum\limits_\alpha^{N_\alpha} \frac{1}{N_i}\sum\limits_i^{N_i} \mathcal{L}\left(\mathrm{data} \left| M_1^{(i)}, M_2^{(i)}, \Lambda^{(\alpha)}(M_1^{(i)}), \Lambda^{(\alpha)}(M_2^{(i)}) \right.\right) \quad \left| \quad \begin{matrix} M_1^{(i)}, M_2^{(i)} \sim P(M_1, M_2|\mathcal{H}) \\ \varepsilon^{(\alpha)} \sim P(\varepsilon|\{\varepsilon\}_A) \end{matrix} \right. ,
\end{equation}
\end{widetext}
where we draw $N_i$ mass realizations for each of the $N_\alpha$ \EOS~realizations.
Within this Monte-Carlo algorithm, we optimize our \KDE~model for $\mathcal{L}(d|\cdots)$ by selecting bandwidths that maximize a cross-validation likelihood based on the public samples (see Appendix~\ref{sec:KDE}).

The overarching composition-marginalized priors are constructed hierarchically, assuming equal prior odds for each composition, which is to say
\begin{multline}
    P(\mathrm{data}|X) = \\
        \frac{1}{3}\big{[} P(\mathrm{data}|X;\mathrm{Hadronic}) \quad \quad \\
            \quad \quad  + P(\mathrm{data}|X;\mathrm{Hyperonic}) \\
                + P(\mathrm{data}|X;\mathrm{Quark}) \big{]}
\end{multline}
for \emph{informed} and \emph{agnostic} priors processes separately.
In the following sections, we analyze GW170817 using both our \emph{agnostic} and \emph{informed} priors.
The full set of results, broken down by composition, is given in Appendix~\ref{sec:supplementary}.

%-------------------------------------------------
\section{Implications for GW170817}
\label{sec:results}

We apply publicly available GW data from GW170817~\cite{GW170817samples} to our priors to infer the system's properties \textit{a posteriori}.
Section~\ref{sec:macroscopic} focuses on macroscopic observables associated with the inspiral stage of the coalescence, such as component masses ($M_{1,2}$), tidal deformabilities ($\Lambda_{1,2}$), and radii ($R_{1,2}$), while Section~\ref{sec:progenitor} focuses on the nature of the progenitor system and the remnant.
Throughout this section, we quote medians and 90\% highest-probability-density credible regions unless otherwise stated.

%------------------------
\subsection{Constraints on GW170817's macroscopic observables}
\label{sec:macroscopic}

\begin{figure*}
    \begin{minipage}{0.49\textwidth}
        \begin{center}
            \emph{model-agnostic} \\
            \includegraphics[width=1.0\textwidth, clip=True, trim=0.5cm 1.9cm 0.5cm 0.4cm]{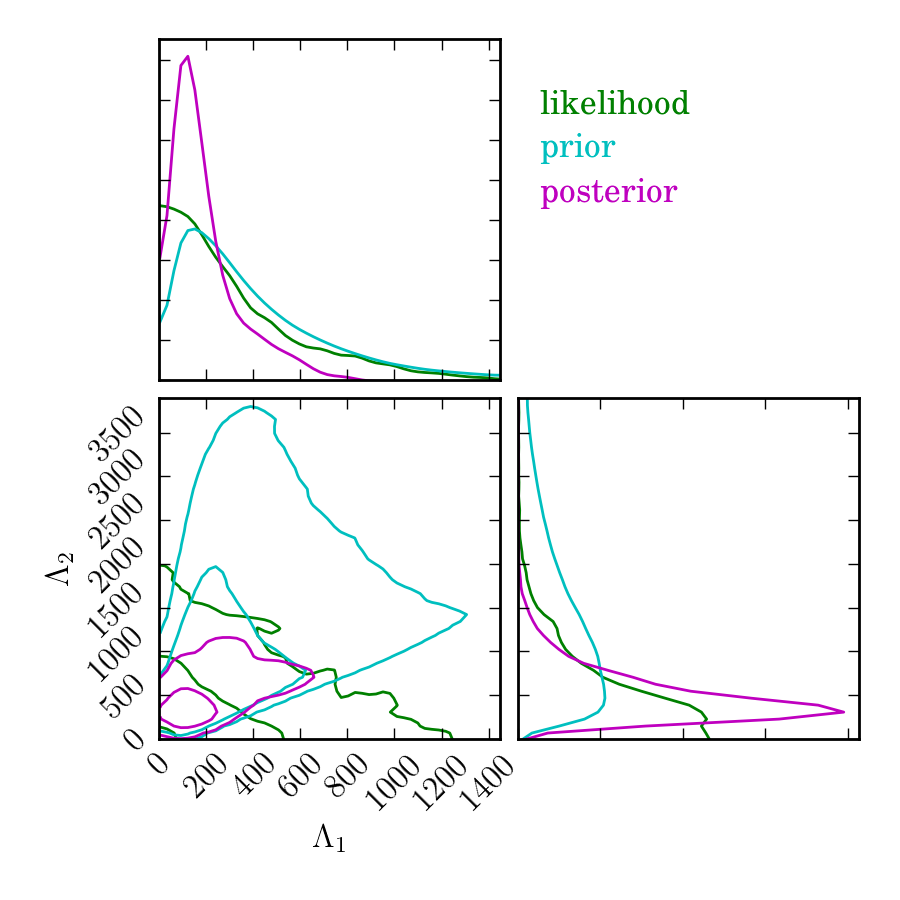} \\
            \includegraphics[width=1.0\textwidth, clip=True, trim=0.5cm 0.6cm 0.5cm 5.0cm]{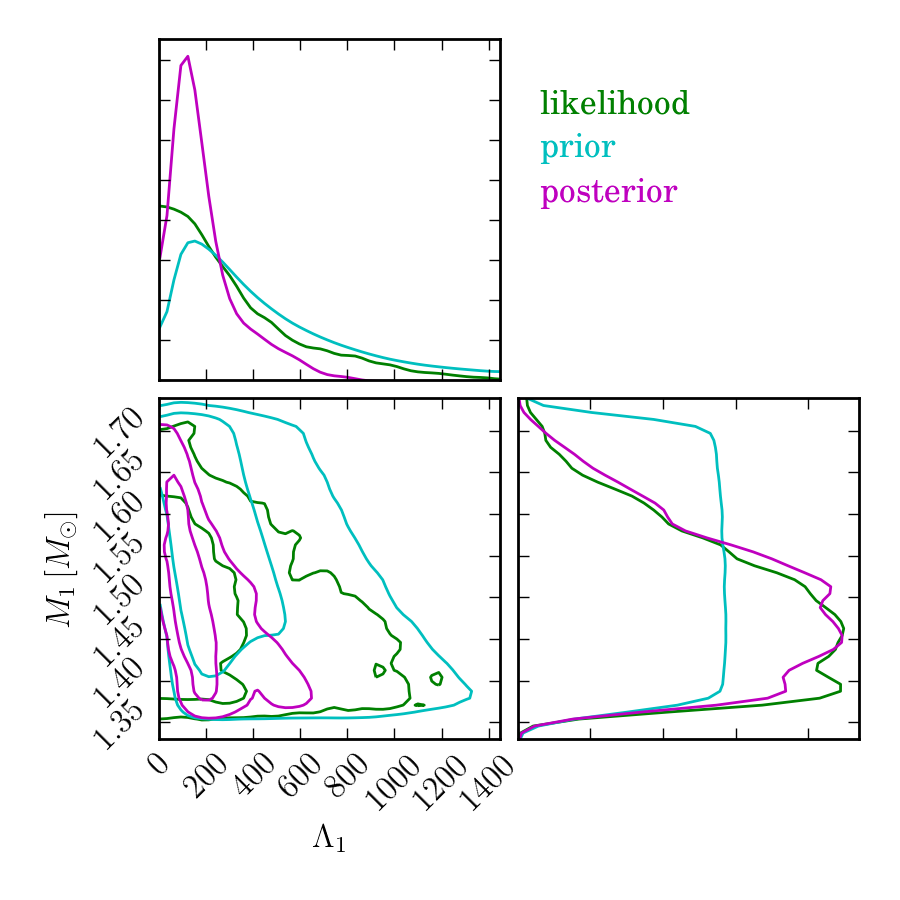}
        \end{center}
    \end{minipage}
    \begin{minipage}{0.49\textwidth}
        \begin{center}
            \emph{model-informed} \\
            \includegraphics[width=1.0\textwidth, clip=True, trim=0.5cm 1.9cm 0.5cm 0.4cm]{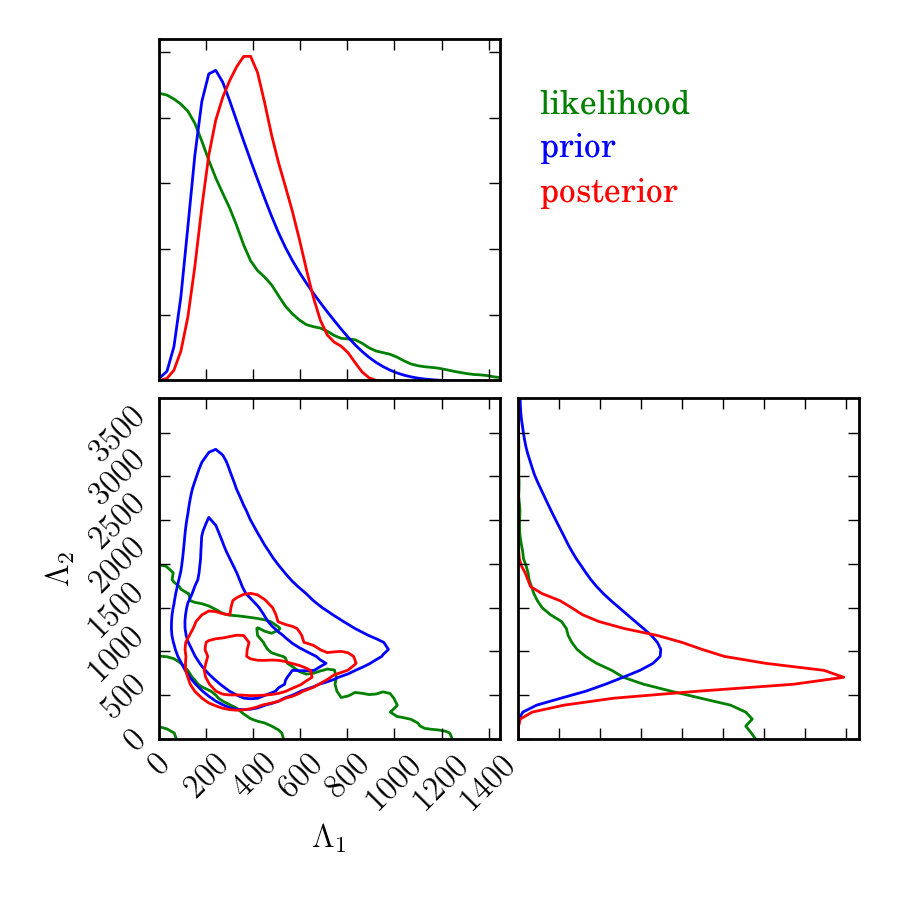} \\
            \includegraphics[width=1.0\textwidth, clip=True, trim=0.5cm 0.6cm 0.5cm 5.0cm]{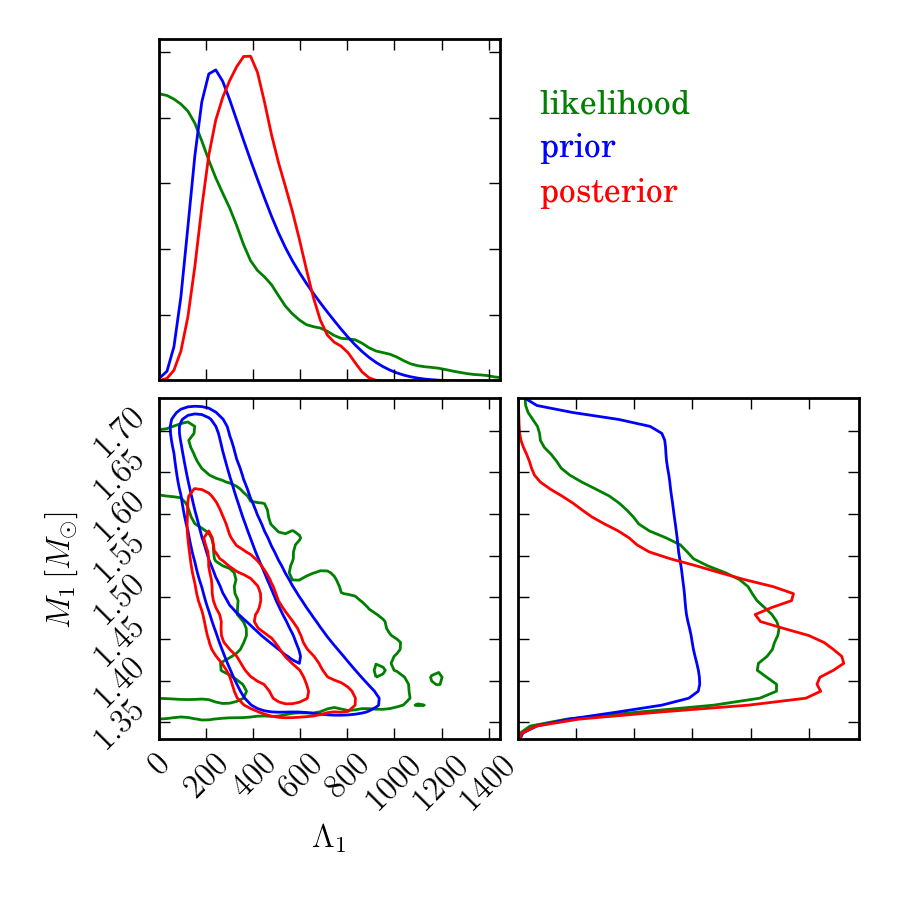}
        \end{center}
    \end{minipage}
    \caption{
        Distributions for $M_1$, $\Lambda_1$, and $\Lambda_2$ after marginalizing over \NS~composition.
        (\emph{Left}) \emph{model-agnostic} prior (\emph{cyan}), posterior (\emph{magenta}), and low-spin marginal likelihood (\emph{green}).
        (\emph{Right}) \emph{model-informed} prior (\emph{blue}), posterior (\emph{red}), and low-spin marginal likelihood (\emph{green}).
        Contours in the joint distributions denote highest-posterior-density 50\% and 90\% credible regions.
    }
    \label{fig:macro marg}
\end{figure*}

We begin with posterior constraints on the macroscopic properties of GW170817, assuming both compact objects were slowly spinning \NSs.
Table~\ref{table:macroscopic} enumerates credible regions for various properties of GW170817's constituents, and Figure~\ref{fig:macro marg} demonstrates the correlations between some of these properties.
In principle, our inference constrains any \EOS-dependent observable associated with the event, but we focus on those that either directly impact the \GW~waveform or have been discussed extensively elsewhere in the literature.
While we show low-dimensional projections of our data, we perform our inference in the four-dimensional space spanned by $M_1$, $M_2$, $\Lambda_1$, and $\Lambda_2$, and therefore posterior constraints may not be intuitive from the the low-dimensional projections in our figures.

We analyze publicly available posterior samples~\footnote{These samples were generated with uniform (flat) priors in $M_1$, $M_2$, $\Lambda_1$, and $\Lambda_2$, and therefore are drawn from a distribution proportional to the likelihood, which is what we model in our analysis.} with component spins constrained to be below $0.05$~\cite{GW170817samples}, and \result{$\gtrsim O(10^{5})$} synthetic \EOSs~drawn from each of our priors with 50 mass realizations per \EOS.
We find that relative uncertainties obtained with the \emph{informed} prior are generally smaller than those with the \emph{agnostic} prior, although the distributions of some parameters, like tidal deformabilities, are centered on larger values with the \emph{informed} prior and can produce larger credible intervals in absolute terms.
The inferred component mass distributions do not depend strongly on the \EOS~prior assumed, although they do change slightly \textit{a posteriori} because they are correlated with the tidal deformability and radius, which are more sensitive to \EOS~assumptions.
Specifically, the \emph{agnostic} prior supports extreme \EOSs~and produces rather low values for $\Lambda$ and $R$ \textit{a posteriori}.
The \emph{informed} prior prefers stiffer \EOSs~and yields correspondingly larger $\Lambda$ and $R$.
Table~\ref{table:macroscopic} lists the precise credible regions obtained.
We note that the inferred tidal deformabilities are consistent with \LandryEssick, and both the $\Lambda$ and $R$ credible regions are consistent with others in the literature.
In particular, Ref.~\cite{GW170817eos} constrained $R_1$ and $R_2$ separately, finding both to be \externalresult{$11.9^{+1.4}_{-1.4}$} km, while Ref.~\cite{De2018} assumed a common radius for the \NSs~and constrained it to \externalresult{8.9--13.2} km.
We find $R_{1}=\ROneAgnostic$ km and $R_2=\RTwoAgnostic$ km with our \emph{agnostic} prior, confirming the validity of the common radius assumption \textit{a posteriori}.
Similarly, Ref.~\cite{GW170817properties} finds \externalresult{$\tilde{\Lambda}=300^{+420}_{-230}$}, while Ref.~\cite{De2018} quotes $\tilde{\Lambda} =  222^{+420}_{-138}$ for a uniform component mass prior.
Our \emph{agnostic} prior yields $\tilde{\Lambda}=\LTildeAgnostic$ \footnote{While our \emph{informed} priors constrain $R$ and $\tilde{\Lambda}$ more tightly, we note that Refs.~\cite{GW170817eos} and~\cite{De2018} are more similar to our \emph{agnostic} prior and therefore consider this the relevant comparison.}.
Finally, our constraints remain marginally consistent with those quoted in Refs.~\cite{Radice2018a, Radice2018b, Margalit2017, Coughlin2018} based on electromagnetic observations, which set a lower bound of \externalresult{$\tilde{\Lambda}\gtrsim300$}.

Like Figure 1 in Ref.~\cite{GW170817eos} and Figure 6 in \LandryEssick, Figure~\ref{fig:macro marg} shows the prior, likelihood, and posterior credible regions for $M_1$, $\Lambda_1$, and $\Lambda_2$ with our composition-marginalized priors.
The preference for softer \EOSs~in the \emph{agnostic} prior, both \textit{a priori} and \textit{a posteriori}, is readily apparent, as is the fact that the \emph{agnostic} prior encompasses a wider range of possibilities than the \emph{informed} prior.

\begin{table*}
    \caption{
        Medians \textit{a posteriori} and highest-probability-density 90\% credible regions for macroscopic observables and central densities associated with GW170817.
    }
    \label{table:macroscopic}
    \begin{tabular}{crrrrrrrcc}
        \hline
        Prior ($\mathcal{H}_i$)     & \multicolumn{1}{c}{$M_1\,[M_\odot]$} & \multicolumn{1}{c}{$M_2\,[M_\odot]$} & \multicolumn{1}{c}{$\Lambda_1$} & \multicolumn{1}{c}{$\Lambda_2$} & \multicolumn{1}{c}{$\tilde\Lambda$} & \multicolumn{1}{c}{$R_1\,[\mathrm{km}]$} & \multicolumn{1}{c}{$R_2\,[\mathrm{km}]$} & \multicolumn{1}{c}{$\rho_{c,1}\,[10^{14}\,\mathrm{g}/\mathrm{cm}^3]$} & \multicolumn{1}{c}{$\rho_{c,2}\,[10^{14}\,\mathrm{g}/\mathrm{cm}^3]$} \\
        \hline
        \hline
        \emph{informed} & \MOneInformed & \MTwoInformed & \LOneInformed & \LTwoInformed & \LTildeInformed & \ROneInformed & \RTwoInformed & \RhocOneInformed & \RhocTwoInformed \\
        \emph{agnostic} & \MOneAgnostic & \MTwoAgnostic & \LOneAgnostic & \LTwoAgnostic & \LTildeAgnostic & \ROneAgnostic & \RTwoAgnostic & \RhocOneAgnostic & \RhocTwoAgnostic \\
        \hline
    \end{tabular}
\end{table*}

Table~\ref{table:macroscopic} also reports credible intervals for the central densities of the two components.
These are consistent with those reported in both \LandryEssick~and Ref.~\cite{GW170817eos}.
We again note the general trend that the \emph{agnostic} prior prefers softer \EOSs~with correspondingly compact stars containing denser cores and higher central pressures.

As discussed in \LandryEssick, we obtain different \textit{a posteriori} credible regions with different priors.
However, neither of our priors are strongly favored by the data, with a Bayes Factor of $B^\mathcal{A}_\mathcal{I}=\BayesAgnInf$ between them.
This constitutes marginal evidence in favor of the \emph{agnostic} priors over the \emph{informed} priors, but serves mainly to demonstrate that posterior constraints on macroscopic observables and the \EOS~need to be interpreted with care, giving consideration to the underlying prior assumptions.

%------------------------
\subsection{Implications for GW170817's progenitor and remnant}
\label{sec:progenitor}

Section~\ref{sec:macroscopic}'s results implicitly assume both constituents are \NSs.
This is reasonable given the masses involved, which are well below the minimum \Mmax~allowed in our priors, and the electromagnetic counterparts observed in coincidence~\cite{GW170817mma, GW170817grb}.
However, we also consider the possibility that either (or both) of the constituents could be a \BH.
Although credible regions similar to Section~\ref{sec:macroscopic} could be derived for each case, we simply focus on the evidence for each progenitor type to determine whether the GW data alone can rule out the presence of at least one \BH~in the system.

Our evidence calculation involves Monte-Carlo integrals over a KDE representation of $\mathcal{L}(d|\cdots)$ along boundaries where $\Lambda_1$, $\Lambda_2 \rightarrow 0$.
Such boundaries can introduce biases in the KDE representation of $\mathcal{L}$.
After optimizing our \KDE, we find that these effects introduce percent-level systematics.

We begin by investigating the nature of the progenitor system: did it consist of two \NSs~(BNS), a \NS~and a lighter \BH~(\NSBH), a \NS~and a heavier \BH~(\BHNS), or two \BHs~(\BBH)?
Table~\ref{table:progenitor model selection} quotes the posterior probability of each progenitor type assuming equal prior odds.
We include results assuming both the low- and high-spin priors from Ref.~\cite{GW170817properties}, although the low-spin prior is motivated by the maximum observed rotation frequencies of pulsars in galactic \BNS~systems that will merge within a Hubble time (J0737--3039A~\cite{Lyne2004} and J1946+2052~\cite{Stovall2018}) and may not be applicable to \BHs.
We also test for the presence of at least one NS.
We find that GW170817 is relatively inconsistent with a \BBH~based on \GW~data alone, disfavored by a factor of $\BayesNSBHHispinAgnostic$ relative to a progenitor with at least one \NS~with the high-spin \emph{agnostic} prior, assuming $P(\BNS)=P(\NSBH)=P(\BHNS)=1/6$ and $P(\BBH)=1/2$.
The weak preference for the \BNS~model with our \emph{agnostic} prior is likely due to the relatively large Occam factor incurred by the extra freedom associated with $\Lambda_{1,2}$ compared to the \BBH~model.
Therefore, even though the maximum likelihood within the \BNS~model is consistently four times larger than in the \BBH~model, the marginal likelihoods are comparable.
This is also true to a lesser extent with the \emph{informed} prior, but its stricter \textit{a priori} assumptions also reduce the maximum likelihood and therefore the marginal evidence for the \BNS~model.
In addition, of the possible progenitors containing a \NS, GW170817 data weakly favor a NSBH, with $B^\NSBH_\BNS(\chi_i\leq0.89,\,\text{\emph{agnostic}})=\BayesNSBHBNSAgnostic$.
This is likely due to marginally better matches to the data when $\Lambda_2\rightarrow0$ compared to the larger $\Lambda(M_2)$ required by our priors, instead of just an Occam factor, since \NSBH~models are preferred over \BHNS~models despite both having approximately equal prior volumes.
We obtain qualitatively similar results with the high- and low-spin priors.

Ref.~\cite{GW170817modelselection} computed similar Bayes factors for individual tabulated \EOSs, finding \externalresult{$\ln B^\BNS_\BBH \lesssim 2$} for their wide mass prior, with the majority of \EOSs~considered yielding \externalresult{$\ln B^\BNS_\BBH\sim0$}, in good agreement with our $\ln B^\BNS_\BBH(\chi_i\leq0.89,\,\text{\emph{agnostic}})=\lnBayesBNSBBHHispinAgnostic$.

Our calculations demonstrate that GW170817 is more consistent with a system containing at least one \NS~as compared to a \BBH.
Of course, this result is unsurprising given that an electromagnetic counterpart was observed.
Given this, we predict the amount of matter available outside the remnant, similar in spirit to Refs.~\cite{Coughlin2018} and \cite{GW170817kilonova}, among others \cite{GW170817mma,Kasen2017,Cowperthwaite2017,Villar2017,GW170817modelselection}.
Using the fitting formula reported in the appendix of Ref.~\cite{Coughlin2018}, which depends primarily on stellar compactness, we estimate the amount of dynamical ejecta and its velocity, finding $\Meject=\MejectAgnostic$ (\MejectInformed) $M_\odot$ and $\Veject=\VejectAgnostic$ (\VejectInformed) with our \emph{agnostic} (\emph{informed}) prior, generally in good agreement with estimates of the contribution of dynamical ejecta in the literature.
We note that our error bars account for the residual \EOS~uncertainty, but modeling systematics associated with Ref.~\cite{Coughlin2018}'s fit surely contribute to the error budget as well.
As has been previously noted, this suggests the dynamical ejecta were only a small part of the total ejecta of $\gtrsim0.05\,M_\odot$ which powered GW170817's kilonova~\cite{Coughlin2018, GW170817kilonova}. 

Numerical relativity simulations typically do not extend far enough past merger to observe mass ejected via disk winds, which are expected to dominate the total ejected mass.
Similarly, it is difficult to estimate the lanthanide fraction, which determines the opacity of the ejected material and the kilonova's color, from first principles based on $M_1$, $M_2$, and the \EOS.
However, were such models available, our posterior processes would immediately bound the expected kilonova properties, just as we already constrain the contribution of dynamical ejecta.

\begin{table*}
    \caption{
        Posterior probabilities for progenitor systems assuming equal prior odds.
        We compare the evidence for a binary \NS~(\BNS), a \NSBH~($M_1$ is a \NS), a \BHNS~($M_1$ is a \BH), and a binary \BH~(\BBH).
        Monte-Carlo sampling uncertainties are approximately three orders of magnitude smaller than the point estimates.
        Reported uncertainties approximate systematic error from our \KDE~model of the \GW~likelihood, which is largest for the \BBH~hypothesis.
    }
    \label{table:progenitor model selection}
    \begin{tabular}{cccccc}
        \hline
        Spin Prior & \EOS~Prior ($\mathcal{H}_i$)     & $P(\BNS|\mathrm{data};\mathcal{H}_i)$ & $P(\BHNS|\mathrm{data};\mathcal{H}_i)$ & $P(\NSBH|\mathrm{data};\mathcal{H}_i)$ & $P(\BBH|\mathrm{data};\mathcal{H}_i)$ \\
        \hline
        \hline
        \multirow{2}{*}{$|\chi_i|\leq0.05$} & \emph{informed} & \PBNSLoSpinInformed & \PBHNSLoSpinInformed & \PNSBHLoSpinInformed & \PBBHLoSpinInformed \\
                                            & \emph{agnostic} & \PBNSLoSpinAgnostic & \PBHNSLoSpinAgnostic & \PNSBHLoSpinAgnostic & \PBBHLoSpinAgnostic \\
        \hline
        \multirow{2}{*}{$|\chi_i|\leq0.89$} & \emph{informed} & \PBNSHiSpinInformed & \PBHNSHiSpinInformed & \PNSBHHiSpinInformed & \PBBHHiSpinInformed \\
                                            & \emph{agnostic} & \PBNSHiSpinAgnostic & \PBHNSHiSpinAgnostic & \PNSBHHiSpinAgnostic & \PBBHHiSpinAgnostic \\
        \hline
    \end{tabular}
\end{table*}

%-------------------------------------------------
\section{Implications for Neutron Star Properties}
\label{sec:ilove}

In addition to examining the inferred properties of GW170817, we can use the GW data to inform our knowledge of \NSs~in general.
Specifically, we compute posterior processes for various functional degrees of freedom, including the \EOS~itself and several derived relations between macroscopic observables.
If all \NSs~share a single universal \EOS, then these results are immediately applicable to other systems.
Tight constraints on these relationships imply consistency tests of the universal-\EOS~hypothesis with observations of other systems~\cite{Landry2018, Kumar2019}.
Tables~\ref{table:ilov central} and~\ref{table:ilov} summarize our conclusions, and we discuss a few salient points in more detail below.
As in Section~\ref{sec:results}, we report medians and 90\% highest-probability-density credible regions unless otherwise noted.

%------------------------
\subsection{Posterior processes for the \EOS}
\label{sec:EOS process}

\begin{figure*}
    \begin{minipage}{0.49\textwidth}
        \begin{center}
            \emph{model-agnostic} \\
            \includegraphics[width=1.0\textwidth]{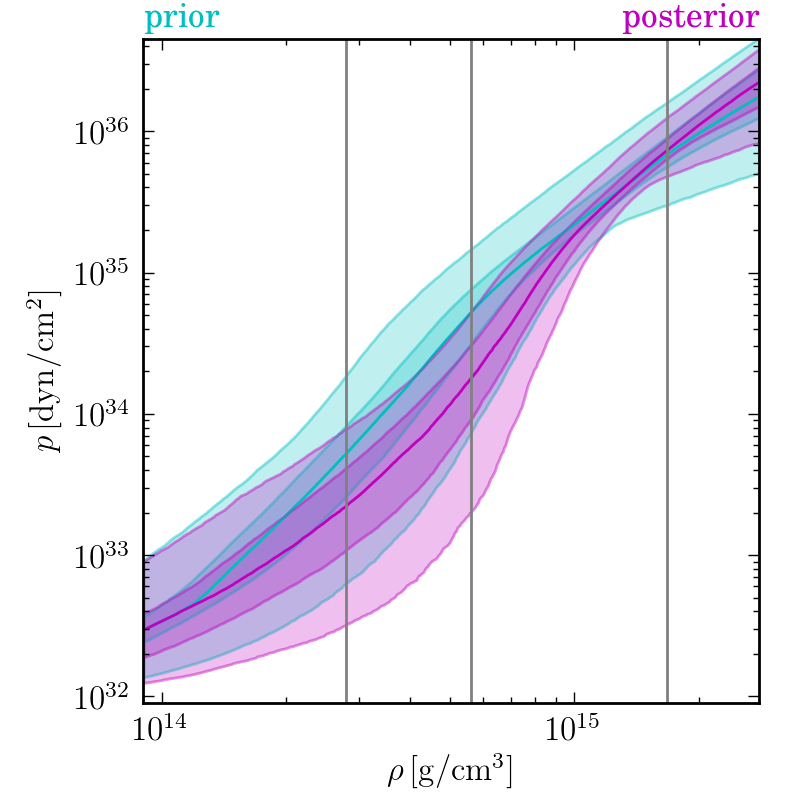}
        \end{center}
    \end{minipage}
    \begin{minipage}{0.49\textwidth}
        \begin{center}
            \emph{model-informed} \\
            \includegraphics[width=1.0\textwidth]{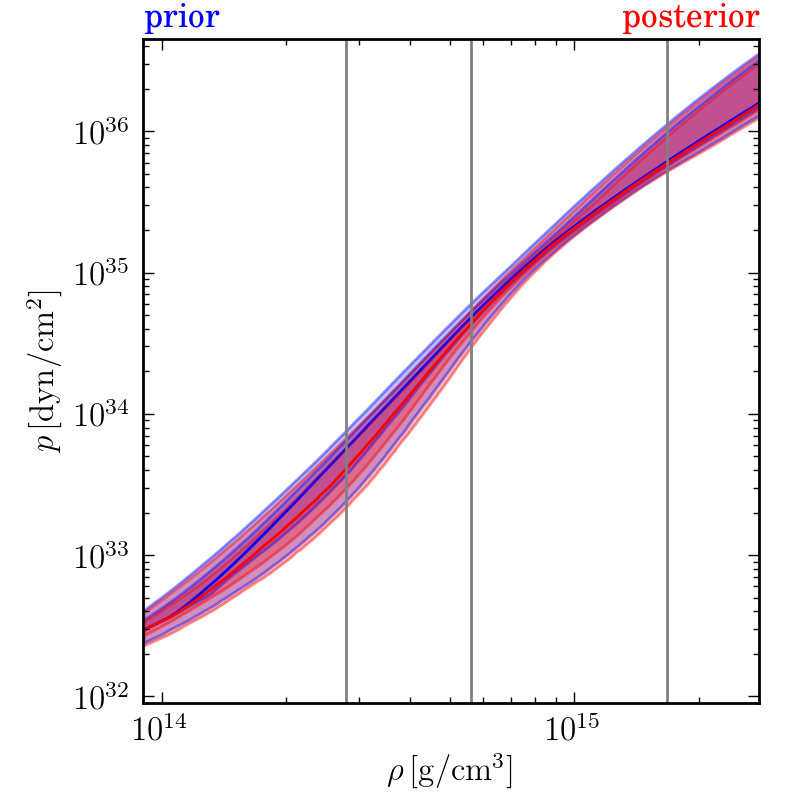}
        \end{center}
    \end{minipage}
    \caption{
        \EOS~processes after marginalizing over composition.
        (\emph{Left}) \emph{model-agnostic} prior (\emph{cyan}) and posterior (\emph{magenta}) processes.
        (\emph{Right}) \emph{model-informed} prior (\emph{blue}) and posterior (\emph{red}) processes.
        Shaded regions correspond to 50\% and 90\% symmetric marginal credible regions for the pressure at each density.
        Solid lines denote the median and vertical lines denote \rhonuc, 2\rhonuc, and 6\rhonuc.
    }
    \label{fig:eos marg}
\end{figure*}

We begin with an inference of the \EOS~itself, as shown in Figure~\ref{fig:eos marg}.
\textit{A posteriori}, we observe a general trend towards lower pressures, particularly between \rhonuc~and 2\rhonuc, with a trend back towards pressures near the \textit{a priori} median at higher densities.
Table~\ref{table:ilov central} quantifies the uncertainty in pressure at a few reference densities.

The constraints we obtain are slightly different than, but consistent with, those reported in \LandryEssick, which is expected from the differences in our \GP~prior processes.
Our \emph{agnostic} and \emph{informed} results bracket those reported in Ref.~\cite{GW170817eos}'s parametric analysis, namely \externalresult{$p(2\rhonuc)=3.5^{+2.7}_{-1.7}\times10^{34}$ dyn/cm$^2$} and \externalresult{$p(6\rhonuc)=9.0^{+7.9}_{-2.6}\times10^{35}$ dyn/cm$^2$}.
Specifically, our \emph{agnostic} results are systematically lower \textit{a posteriori} than Ref.~\cite{GW170817eos}'s, while our \emph{informed} pressure bounds are centered above Ref.~\cite{GW170817eos} at 2\rhonuc.
At 6\rhonuc, however, our \emph{informed} process lies below both Ref.~\cite{GW170817eos} and our \emph{agnostic} results, although the uncertainties are broad.
This inversion likely occurs because GW170817 has little constraining power at high densities, but the \emph{informed} results favor the presence of quark matter, which systematically softens \textit{a priori} in this regime.

The trend toward low pressures between \rhonuc~and 2\rhonuc~is likely driven by at least two factors.
First, the \NS~radius and tidal deformability are known to correlate strongly with the pressure in that region~\cite{Lattimer2001, Read2009}, and therefore GW170817's preference for small $\Lambda$ manifests as a preference for lower pressures in this region.
However, GW170817's component masses likely have central densities below $10^{15}\,\mathrm{g}/\mathrm{cm}^3$, and therefore the GW data only contains information about the \EOS~at densities lower than this.
At significantly higher densities, then, the \EOS~posterior will tend to snap back towards the prior.
That tendency is compounded by the requirement that all \EOSs~support $1.93\,M_\odot$ stars, which forces initially soft \EOSs~to stiffen at higher densities in order to support massive stars.

%------------------------
\subsection{Posterior distributions and processes for macroscopic observables}
\label{sec:macro process}

\begin{figure*}
    \begin{center}
        \emph{model-agnostic} \\
    \end{center}
    \begin{minipage}{0.45\textwidth}
        \begin{center}
            \includegraphics[width=1.0\textwidth, clip=True, trim=0.0cm 1.2cm 0.0cm 0.50cm]{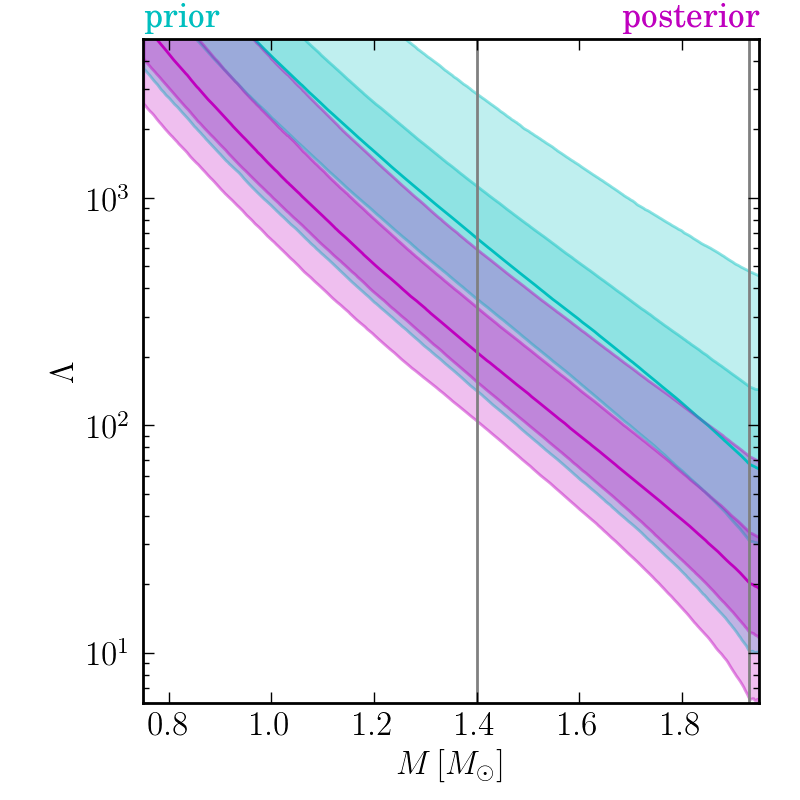} \\
            \includegraphics[width=1.0\textwidth, clip=True, trim=0.0cm 1.2cm 0.0cm 0.45cm]{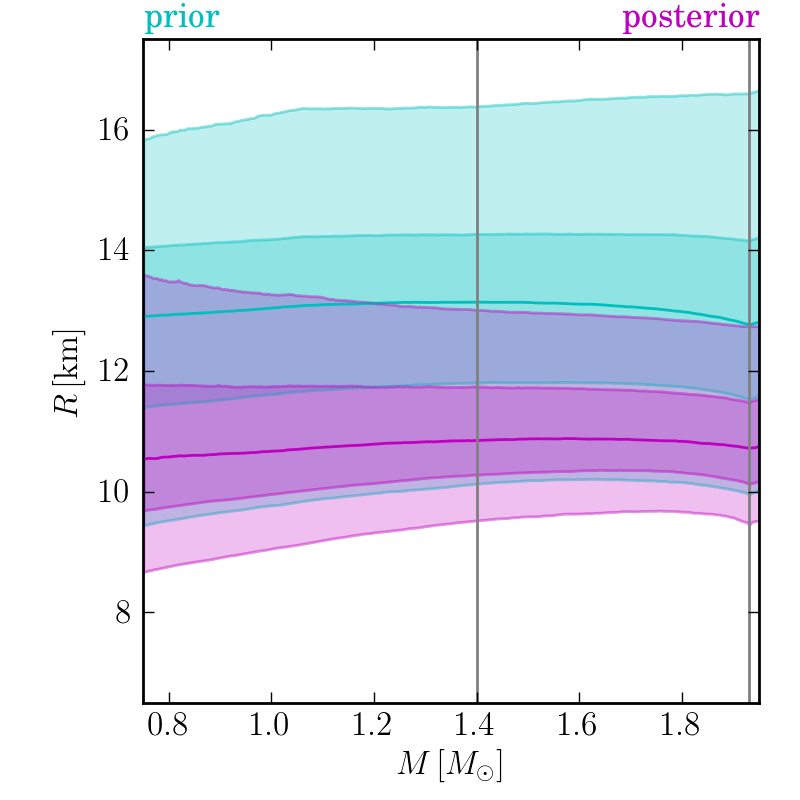} \\
            \includegraphics[width=1.0\textwidth, clip=True, trim=0.0cm 0.0cm 0.0cm 0.45cm]{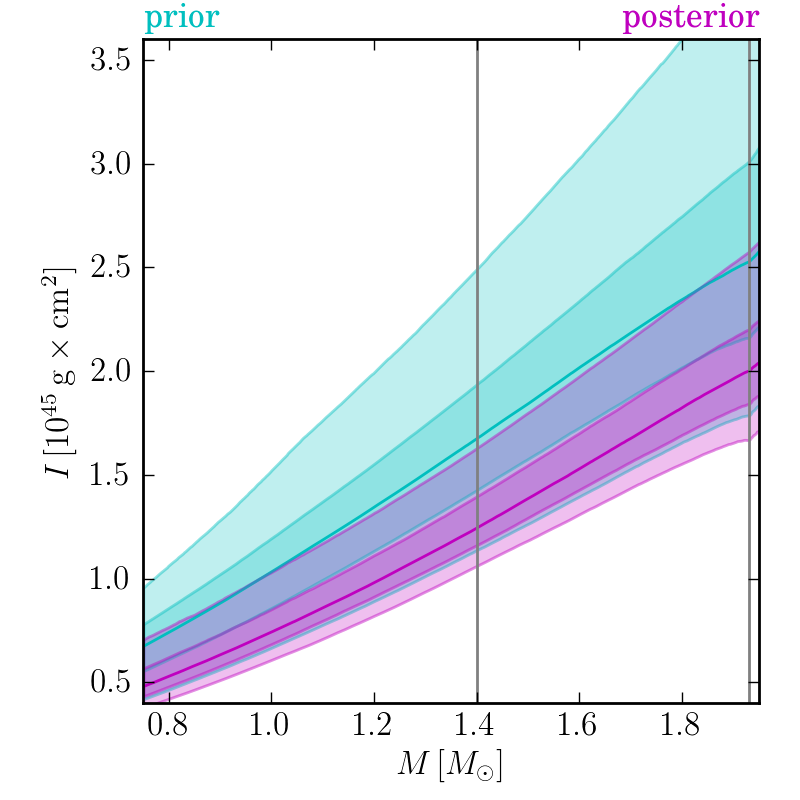}
        \end{center}
    \end{minipage}
    \begin{minipage}{0.45\textwidth}
        \begin{center}
            \includegraphics[width=1.0\textwidth, clip=True, trim=0.0cm 1.2cm 0.0cm 0.50cm]{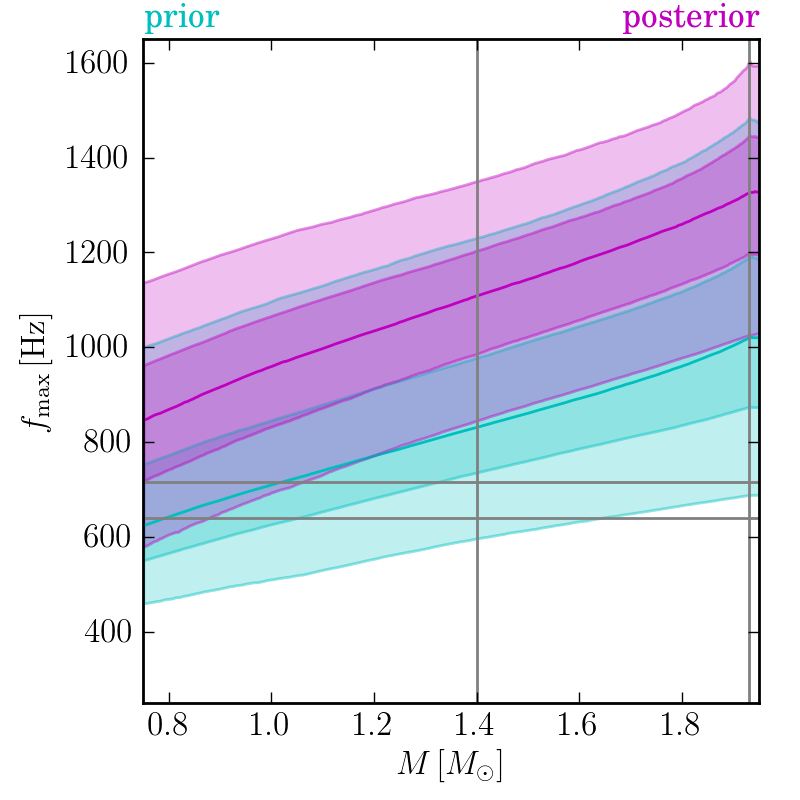} \\
            \includegraphics[width=1.0\textwidth, clip=True, trim=0.0cm 1.2cm 0.0cm 0.45cm]{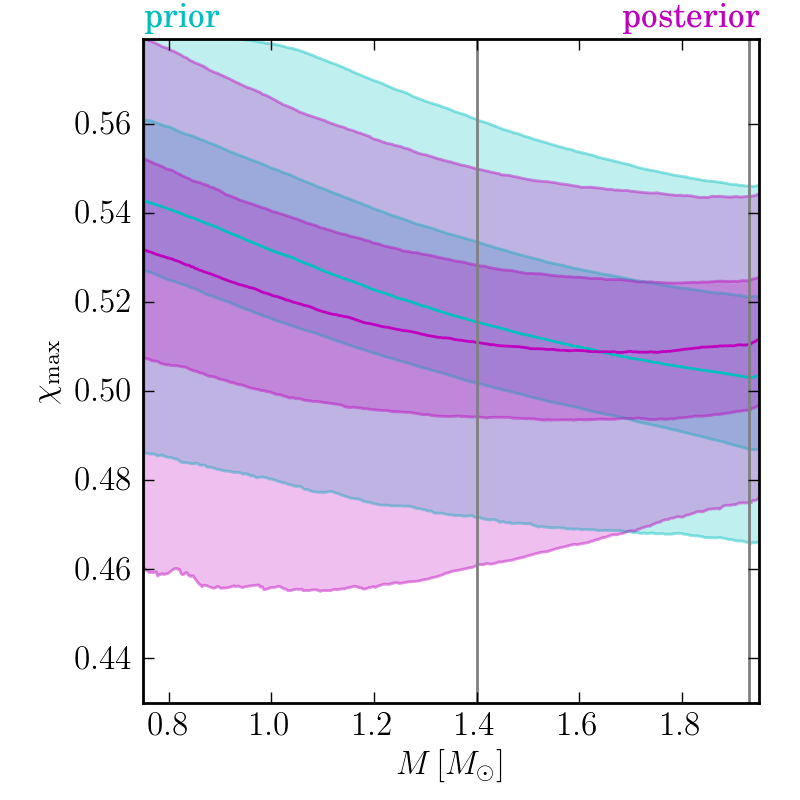} \\
            \includegraphics[width=1.0\textwidth, clip=True, trim=0.0cm 0.0cm 0.0cm 0.45cm]{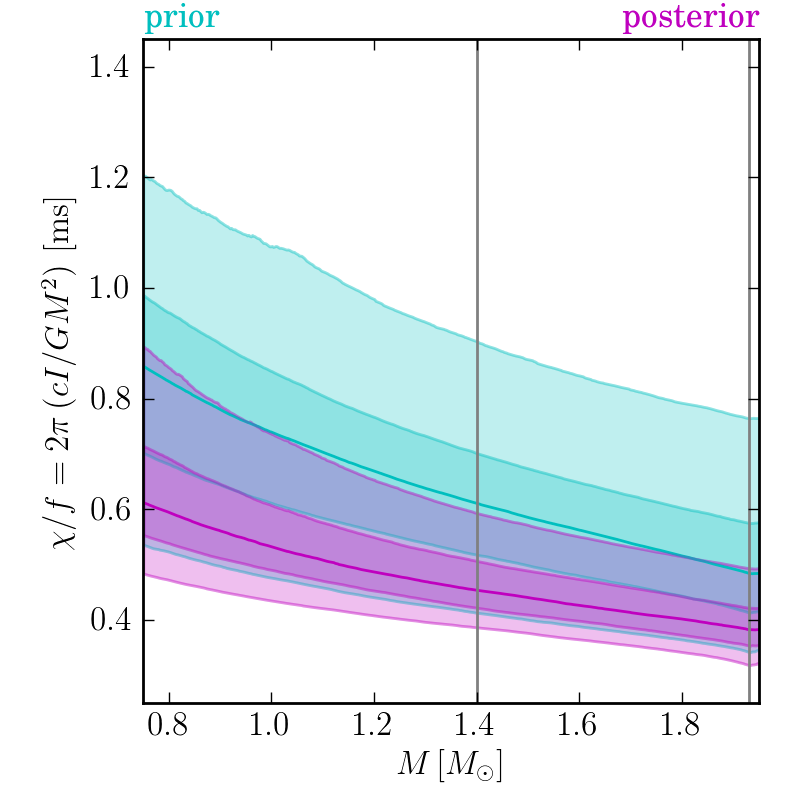}
        \end{center}
    \end{minipage}
    \caption{
        Processes relating a few macroscopic observables after marginalizing over \EOS~composition with the \emph{agnostic} prior.
        Prior (\emph{cyan}) and posterior (\emph{magenta}) processes for (\emph{left}) tidal deformability ($\Lambda$; \emph{top}), radius ($R$; \emph{middle}), and moment of inertia ($I$; \emph{bottom}) as functions of mass as well as (\emph{right}) the maximum spin frequency (\Fbrk; \emph{top}), maximum dimensionless spin parameter (\Chibrk; \emph{middle}), and the dimensionless spin parameter divided by the spin frequency (\emph{bottom}), useful when estimating $\chi$ for a pulsar with unknown mass.
        Shaded regions denote the 50\% and 90\% symmetric credible regions for the marginal distribution of each observable at each mass.
        Solid lines denote the median and vertical lines denote canonical $1.4\,M_\odot$ stars.
        Horizontal grey lines in the \emph{top}-\emph{right} panel denote the measured spin frequencies of J1748--2446ad (\externalresult{$716\,\mathrm{Hz}$}~\cite{Hessels2006}) and B1937+241 (\externalresult{641 Hz}~\cite{Backer1982}), which lie below the 90\% lower limits for \Fbrk~only when $M\gtrsim M_\odot$.
    }
    \label{fig:process marg}
\end{figure*}

\begin{table*}
    \caption{
        Medians \textit{a posteriori} and highest-probability-density 90\% credible regions for a canonical $1.4 \, M_{\odot}$ \NS's central density, and for pressures at several reference densities.
    }
    \label{table:ilov central}
    \begin{tabular}{ccccc}
        \hline
        Prior ($\mathcal{H}_i$)     & \multicolumn{1}{c}{$\rho_{c,1.4}\,[\mathrm{g}/\mathrm{cm}^3]$} & \multicolumn{1}{c}{$p(\rhonuc)\,[\mathrm{dyn}/\mathrm{cm}^2]$} & \multicolumn{1}{c}{$p(2\rhonuc)\,[\mathrm{dyn}/\mathrm{cm}^2]$} & \multicolumn{1}{c}{$p(6\rhonuc)\,[\mathrm{dyn}/\mathrm{cm}^2]$} \\
        \hline
        \hline
        \emph{informed} & \CanonicalRhocInformed & \POneRhonucInformed & \PTwoRhonucInformed & \PSixRhonucInformed \\
        \emph{agnostic} & \CanonicalRhocAgnostic & \POneRhonucAgnostic & \PTwoRhonucAgnostic & \PSixRhonucAgnostic \\
        \hline
    \end{tabular}
\end{table*}

\begin{table*}
    \caption{
        Medians \textit{a posteriori} and highest-probability-density 90\% credible regions for a few canonical macroscopic quantities.
    }
    \label{table:ilov}
    \begin{tabular}{crrcrr}
        \hline
        Prior ($\mathcal{H}_i$) & \multicolumn{1}{c}{$\Lambda_{1.4}$} & \multicolumn{1}{c}{$R_{1.4}\,[\mathrm{km}]$} & \multicolumn{1}{c}{$I_{1.4}\,[10^{45}\,\mathrm{g}\,\mathrm{cm}^2]$} & \multicolumn{1}{c}{$M_{b,1.4}\,[M_\odot]$} & \multicolumn{1}{c}{$\Mmax\,[M_\odot]$} \\
        \hline
        \hline
        \emph{informed}         & \CanonicalLInformed & \CanonicalRInformed & \CanonicalIInformed & \CanonicalMbInformed & \MmaxInformed \\
        \emph{agnostic}         & \CanonicalLAgnostic & \CanonicalRAgnostic & \CanonicalIAgnostic & \CanonicalMbAgnostic & \MmaxAgnostic \\
        \hline
    \end{tabular}
\end{table*}

%-----------
\subsubsection{Maximum mass and binding energy}
\label{sec:maximum mass}

Our posterior process for the \EOS~immediately yields a posterior distribution for the maximum mass of a nonrotating \NS~(\Mmax, sometimes called $M_\mathrm{TOV}$).
Like \LandryEssick, we find smaller \Mmax~are preferred \textit{a posteriori}, and that the shift is larger from prior to posterior for the \emph{agnostic} result.
This is consistent with the preference for softer \EOSs, many of which struggle to support the 1.93$\,M_\odot$ stars required \textit{a priori} and therefore reach their \Mmax~soon thereafter.
We find $\Mmax=\MmaxAgnostic$ (\MmaxInformed) $M_\odot$ with the \emph{agnostic} (\emph{informed}) prior, broadly consistent with other constraints in the literature from both GW and EM data (e.g., Refs.~\cite{Radice2018a, Radice2018b, Margalit2017, Coughlin2018}).

Such constraints may be put to the test as more \NSs~ with large masses are discovered~\cite{Cromartie2019}.
They could play an important role in the analysis of \BHNS~systems for which $\tilde{\Lambda}\sim(M_2/M_1)^4\Lambda_2$ can be quite small.
Indeed, precise knowledge of \Mmax~may be the best way to rule out the possibility that the lighter component is a \NS, particularly if the source is distant enough that electromagnetic counterparts are unlikely to be detectable.
It is worth noting that, while our \emph{agnostic} posterior favors softer \EOSs~in the density regime relevant for GW170817, it produces looser \Mmax~constraints than the \emph{informed} posterior, supporting larger \Mmax~than are allowed by the \emph{informed} prior.
This is associated with the additional model freedom within the \emph{agnostic} prior, which allows the \EOSs~to vary significantly at densities larger than those occurring within GW170817's components.

Table~\ref{table:ilov} also reports the baryonic mass $M_{b,1.4}$ of a canonical $1.4\,M_\odot$ \NS.
Besides its possible relevance for the amount of matter available to power electromagnetic counterparts, the difference between $M_b$ and the \NS~mass defines the star's binding energy, with more compact stars corresponding to larger $M_b$ at fixed $M$.
GW170817 suggests that canonical \NSs~typically have binding energies of \result{0.15--0.2$\,M_\odot$}, corresponding to \result{$>10\%$} of the rest mass of their baryonic content.

%-----------
\subsubsection{Mass-tidal deformability, mass-radius, and mass-moment of inertia relations}
\label{sec:mass-lambda}

Figure~\ref{fig:process marg} shows our posterior processes for several macroscopic observables as a function of the \NS~mass.
We generally find an \textit{a posteriori} preference for smaller $\Lambda$, $R$, and $I$ at a given $M$, consistent with relatively compact \NSs.
This preference is stronger with the \emph{agnostic} posterior than the \emph{informed}, again reflecting the \emph{agnostic} result's preference for particularly soft \EOSs.
Table~\ref{table:ilov} quotes credible regions for a canonical $1.4\,M_\odot$ \NS, showing good agreement with values reported elsewhere.
\LandryEssick~bounded $\Lambda_{1.4}$ to lie between \externalresult{47 and 608} (\externalresult{384 and 719}) at 90\% confidence whith their \emph{agnostic} (\emph{informed}) prior while Ref.~\cite{GW170817eos} found \externalresult{$\Lambda_{1.4}=190^{+390}_{-120}$}.
We find $\Lambda_{1.4}=\CanonicalLAgnostic$ (\CanonicalLInformed).

Ref.~\cite{Capano2019} recently claimed \externalresult{$R_{1.4}=11.0^{+0.9}_{-0.6}$ km} based on GW170817 with a strongly theory-informed prior and a parameterization of the sound-speed at high densities~\cite{Tews2018a, Tews2018b}.
We note that their point-estimate is slightly larger than our \emph{agnostic} result ($\CanonicalRAgnostic\,$km), but that their uncertainties are similar to our \emph{informed} prior.
Their result, then, is likely dependent on their \textit{a priori} assumptions about the \EOS, much like how our \emph{informed} prior strongly influences the posterior constraints.

The derived relations as functions of mass, informed by GW170817, are likely to be of greatest relevance for future observations of both \GW~events and electromagnetic sources.
For example, X-ray timing of pulsars is expected to constrain their masses and radii \cite{Gendreau2012}, radio observations of binary pulsars may measure \NSs' moments of inertia~\cite{Dewdney, Landry2018, Kumar2019}, and additional \GW~observations of coalescing binary \NSs~should produce $M$--$\Lambda$ measurements consistent with the current constraints~\cite{Lackey2015, DelPozzo2013, Agathos2015}.
These measurements will effectively act as a null-test of the hypothesis that all \NSs~share a single \EOS, which is currently difficult to constrain with GW170817 alone (see Ref.~\cite{GW170817properties} and errata of Ref.~\cite{De2018}). 
Indeed, the extreme model freedom allowed by nonparametric analyses will enable novel consistency tests and alternative hypotheses to compare against the universal-\EOS~assumption.

%-----------
\subsection{Maximum spin and asteroseismology}
\label{sec:mass-maximum spin and asteroseismology}

The \EOS~constraints derived from GW170817 have implications for the maximum \NS~spin, since the Keplerian breakup frequency $\Fdyn = \sqrt{GM/R^3}/2\pi$ is sensitive to the radius.
Numerical relativity simulations of rapidly rotating \NSs~show that the maximum spin is typically $\Fbrk\sim0.58\Fdyn$, accurate to $\sim7\%$, after accounting for spin-induced oblateness \cite{Cook1994, Haensel1995, Stergioulas2003, Lattimer2004}.
Here we report estimates for $\Fbrk$, and the corresponding maximum dimensionless spin $\Chibrk = cI(2\pi \Fbrk)/GM^2$, as a function of mass.
We note that our calculation for $I$ assumes slowly rotating stars.
Oblate, rapidly rotating stars will have larger $I$, and therefore our $\Chibrk$ should be interpreted as a lower limit.
Previous studies~\cite{Chakrabarty2003, Chakrabarty2005} noted that the maximum spin obtainable for any \NS~mass is significantly larger than the maximum observed spin frequency, currently \externalresult{$716\,\mathrm{Hz}$}~\cite{Hessels2006}.
We find consistent results, with \result{$\max_M\{\Fbrk\}\gtrsim1.4\,\mathrm{kHz}$}.
However, the maximum spin frequency at a particular mass can be significantly lower, perhaps by as much as a factor of two.
What's more, the proximity of our lower bound on \Fbrk to the observed 716-Hz spin frequency for J1748--2446ad~\cite{Hessels2006} may call into question the need for additional braking mechanisms \cite{Bildsten,Cutler,AnderssonKokkotas,Levin} to limit the spin frequency of recycled millisecond pulsars.

\begin{table*}
    \caption{
        Inferred dimensionless spins for several pulsars with known masses with our \emph{agnostic} (\emph{informed}) composition-marginalized priors.
        The upper limits for J0737$-$3039A, in particular, support the low-spin priors assumed in this work and Refs.~\cite{GW170817discovery, GW170817properties, GW170817eos, GW170817modelselection, Landry2019}.
    }
    \label{tab:pulsars}
    \begin{tabular}{ccrcc}
    \hline
                  \multirow{2}{*}{PSR}   & \multirow{2}{*}{$M\,[M_\odot]$} & \multirow{2}{*}{$f\,[\mathrm{Hz}]$} & \multicolumn{2}{c}{$\chi$} \\
                     &                                 &                                     & \emph{agnostic} & \emph{informed} \\
    \hline
    \hline
        J1807--2500B~\cite{Lynch2012}    & \externalresult{1.37} & \externalresult{238.88} & \result{$0.1101^{+0.0317}_{-0.0187}$} & \result{$0.1319^{+0.0100}_{-0.0118}$} \\
        J0737--3039A~\cite{Lyne2004}     & \externalresult{1.34} & \externalresult{44.05}  & \result{$0.0205^{+0.0059}_{-0.0036}$} & \result{$0.0246^{+0.0019}_{-0.0022}$} \\
        J0348+0432~\cite{Antoniadis2013} & \externalresult{2.01} & \externalresult{25.56}  & \result{$0.0098^{+0.0023}_{-0.0017}$} & \result{$0.0107^{+0.0011}_{-0.0010}$} \\
    \hline
    \end{tabular}
\end{table*}

The corresponding constraints on the maximum dimensionless spin (\Chibrk) demonstrate that \NS~spins must be \result{$\lesssim0.5$} for astrophysically plausible masses.
This provides a natural upper bound on the \NS~spin prior for future Bayesian analyses if the observed distribution of spins in galactic binaries is not applicable to the broader population.
Figure~\ref{fig:process marg} also shows $\chi/f \equiv 2\pi c I/GM^2$ as a function of mass, from which we can compute the dimensionless spin of any pulsar given its observed rotation frequency, even if its mass is not precisely known, with the same caveats as $\Chibrk$ about rapid rotation.
We do this for several pulsars with well-measured masses in Table~\ref{tab:pulsars}.
In particular, the low-spin priors assumed in our work, as well as in Refs.~\cite{GW170817discovery, GW170817properties, GW170817eos, GW170817modelselection}, are motivated by J0737$-$3039A and J1946+2052, with claims that their spins at merger would be below 0.04 and 0.05 \cite{GW170817properties}, respectively.
Our results, which assume $\chi\leq0.05$ \textit{a priori}, support this, with J0737--3039A's current spin inferred to be \result{$0.021^{+0.006}_{-0.004}$} (\result{$0.025^{+0.002}_{-0.002}$}) with our \emph{agnostic} (\emph{informed}) priors.
This is consistent with the dimensionless spin of $\chi \leq 0.034$ inferred for J0737--3039A via universal relations in Ref.~\cite{Landry2018} without the low-spin assumption.

Similarly, we also find the spin of J1807--2500B ($f=238\,\mathrm{Hz}$~\cite{Lynch2012}), one of the fastest pulsars with a well-measured mass, to be \result{$0.11^{+0.03}_{-0.02}$} (\result{$0.13^{+0.01}_{-0.01}$}). Although the dimensionless spin of the fastest known pulsar (J1748--2446ad) depends on its unknown mass, we find \result{$0.25\leq\chi\leq0.65$} for a wide, astrophysically plausible mass range.
In fact, J1748--2446ad's spin frequency is consistent with \result{$f\gtrsim\Fbrk/2$} at 90\% confidence, regardless of mass, and is consistent with \result{$f=\Fbrk$} at $>90\%$ confidence with our \emph{agnostic} prior if \result{$M\lesssim 1.05\,M_\odot$}.

Although beyond the scope of the current work, we also note that precise knowledge of the \EOS~determines the behavior of several dynamical tidal effects.
The \EOS~determines the eigenmode spectra within a \NS, and therefore our posterior processes could be used to determine the exact placement and impact of linear resonant dynamical tidal effects due to $f$-modes and low-order $g$-modes during \GW-driven inspirals (e.g., Refs.~\cite{Lai1994,Yu2016, Yu2017, Xu2017}).
Similarly, knowledge of the $r$-mode spectra could inform the CFS instabilities relevant for millisecond pulsars~\cite{Arras2003, Chakrabarty2003, Arras2005}, and knowledge of the $p$- and $g$-mode spectra could improve models of non-linear, non-resonant secular fluid instabilities relevant during the \GW~inspiral~\cite{Weinberg2013, Venumadhav2014, Weinberg2016, Essick2016, GW170817nltides}.
The precise impact of these last two phenomena, however, also depends strongly on the instabilities' saturation, which themselves are highly uncertain and may prevent precise \EOS~constraints from making strong predictions about their impact on \GW~signals.

%------------------------
\section{Implications for neutron-star composition}
\label{sec:composition model selection}

Finally, we turn to GW170817's implications for \NS~composition.
Unlike previous sections, here we break down our \emph{agnostic} and \emph{informed} priors according to the composition of the \EOSs~upon which they were conditioned, presenting results separately for hadronic, hyperonic and quark GPs.
More results for each composition are available in Appendix~\ref{sec:supplementary}.

To begin, we compare the evidence for each composition assuming both components were slowly spinning \NSs.
Table~\ref{table:composition model selection} shows the posterior probabilities assuming equal prior odds.
Notably, we find weak, but suggestive, evidence in favor of quark matter within \NSs~with the \emph{informed} prior, although the \emph{agnostic} prior prefers \EOSs~containing only hadrons by a similar amount.
The relevance of hadronic vs. quark composition is less clear in the \emph{agnostic} priors by design, though, as they resemble the input \EOSs~less closely.
This is likely just a statement that the tabulated \EOS~from the literature containing quark matter are softer, on average, than those labeled either hadronic or hyperonic.
It is also worth repeating that none of the compositions are overwhelmingly favored.
Nonetheless, the preference for quark \EOSs~is tantalizing, as theoretical considerations suggest there should be a phase transition to quark matter at sufficiently high densities~\cite{Baym2018}.

Regardless of the precise details of \NS~composition, another interesting question is whether there are strong first-order phase transitions within the \EOS, leading to, e.g., distinct hadronic and quark phases of matter.
One possible signature of such strong phase transitions is the existence of a disconnected hybrid star branch in the $M$-$R$ relation.
Stable sequences of \NSs~exist between critical points in the $M$-$R$ relation; the first stable sequence is called the \NS~branch, and any subsequent branches at densities above the phase-transition onset are called hybrid star branches (see, e.g., Ref.~\cite{Schertler}).
Although the presence of only a single stable branch does not preclude the existence of phase transitions, multiple stable branches in the $M$-$R$ relation constitute support for a strong first-order phase transition in the \EOS.
We compute Bayes factors comparing the evidence for \EOSs~that support multiple stable branches at central densities above $0.8\rhonuc$ to those with only a single stable branch above $0.8\rhonuc$ ($B^{n>1}_{n=1}$), finding weak evidence that favors multiple stable branches by a factor of $\BayesBranchesApprox$ with our \emph{agnostic} priors compared to the preference with pulsar data alone.
Table~\ref{table:composition model selection} shows the results for composition-marginalized priors, and the evidence ratios for each composition separately are of the same order of magnitude.
This is far from conclusive, but is suggestive of new physics within \NS~cores.

\begin{table*}
    \caption{
        Posterior probabilities for each composition assuming both components were slowly rotating \NSs~and equal prior odds, as well as Bayes factors for the number of stable branches in the mass-radius relation with the composition-marginalized priors.
        Monte-Carlo sampling uncertainties are approximately two orders of magnitude smaller than the point estimates. The Bayes factor for the \emph{informed} prior is unresolved because its standard deviation is much larger than the point-estimate.
    }
    \label{table:composition model selection}
    \begin{tabular}{ccccc}
        \hline
        Prior ($\mathcal{H}_i$) & $P(\mathrm{Hadronic}|\mathrm{data})$ & $P(\mathrm{Hyperonic}|\mathrm{data})$ & $P(\mathrm{Quark}|\mathrm{data})$ & $B^{n>1}_{n=1}|\textrm{Marginalized}$ \\
        \hline
        \hline
        \emph{informed}         & \PHadInformed & \PHypInformed & \PQrkInformed & \BayesBranchesMrgInformed \\
        \emph{agnostic}         & \PHadAgnostic & \PHypAgnostic & \PQrkAgnostic & \BayesBranchesMrgAgnostic \\
        \hline
    \end{tabular}
\end{table*}

\begin{figure}
    \begin{center}
        \includegraphics[width=1.0\columnwidth, clip=True, trim=0.0cm 0.7cm 0.0cm 0.5cm]{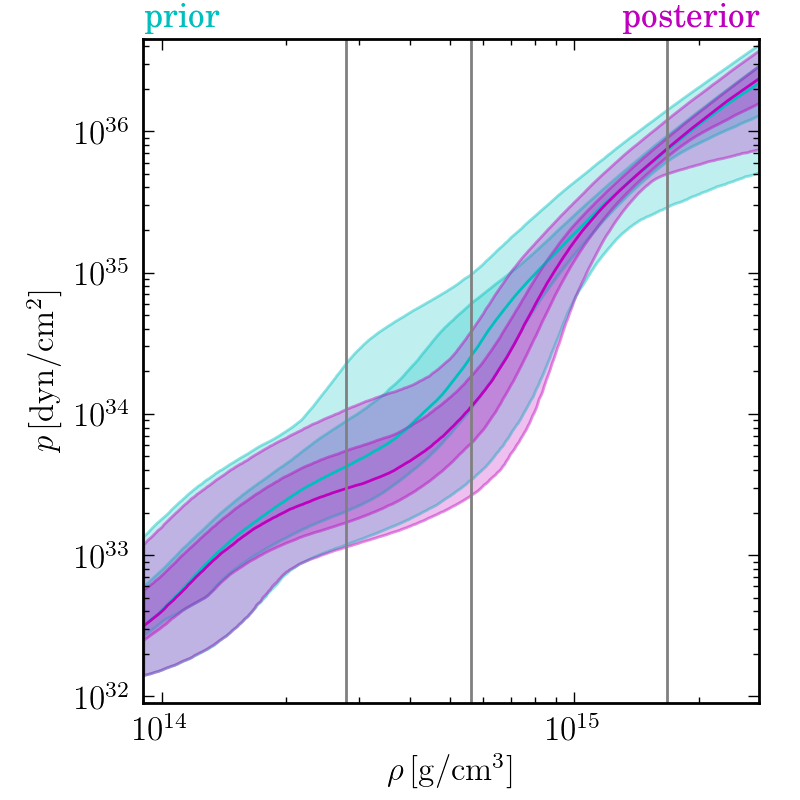} \\
    \end{center}
    \caption{
        \emph{Agnostic} \EOS~prior (\emph{cyan}) and posterior (\emph{magenta}) processes for \EOSs~that support multiple stable branches in the $M$-$R$ relation above $\rho_c=0.8\rhonuc$ after marginalizing over composition.
        The equivalent processes for \EOSs~that support a single stable branch are indistinguishable from Figure~\ref{fig:eos marg}.
        The general preference for softer \EOSs~\textit{a posteriori} manifests as a dramatic softening at or below \rhonuc~before stiffening at approximately 2$\rhonuc$.
        \emph{Gray} lines denote \rhonuc, 2\rhonuc, and 6\rhonuc, respectively.
        Bayes factors between multiple and single stable branches for the agnostic priors are $B^{n>1}_{n=1}=\BayesBranchesMrgAgnostic$, \BayesBranchesHadAgnostic, \BayesBranchesHypAgnostic, and \BayesBranchesQrkAgnostic~for the marginalized, hadronic, hyperonic, and quark compositions, respectively.
    }
    \label{fig:eos given n>1}
\end{figure}

Pursuing this further, Figure~\ref{fig:eos given n>1} shows our \emph{agnostic} prior and posterior processes conditioned on the number of stable branches above $0.8\rhonuc$.
From this we see that, assuming the \EOS~supports at least one disconnected hybrid branch, GW170817 noticeably prefers \EOSs~that dramatically soften near \rhonuc~before stiffening significantly around $2\rhonuc$.
While we expect this type of behavior within \EOSs~that have multiple stable branches, our priors do not have any particular preference for the phase transition to occur in this density range.
Intriguingly, the central densities inferred for GW170817 (see Table~\ref{table:macroscopic}) suggest that any exotic particles associated with the putative phase transition around $\rhonuc$ would have been present within GW170817's components' cores before they coalesced. This finding is consistent with Ref.~\cite{Paschalidis2018}'s conclusion that tidal deformability constraints from GW170817 cannot rule out the presence of a hybrid star.
Figure~\ref{fig:eos marg} shows the processes regardless of the number of stable branches and is dominated by \EOSs~with a single stable branch since these are favored \emph{a priori} by a factor of $\sim10$.
While we see the same general trend toward softer \EOSs, this manifests as a general decrease in pressure at all densities, whereas there is a notable preference for softening and stiffening at specific densities when there are multiple stable branches.

We also note that the posterior preference for multiple stable branches depends on the precise lower limit of \Mmax~allowed in our priors.
The requirement that $\Mmax\geq1.93\,M_\odot$ forces the \EOS~to become stiff at high densities, thereby imparting the preference for \EOSs~that stiffen again after they initially soften.
Without that requirement, \EOSs~that do not stiffen significantly (and hence support only a single stable branch) can still reproduce the GW170817 data reasonably well, weakening the modest preference for \EOSs~with multiple stable branches.
We expect observations of more massive pulsars (e.g., \cite{Cromartie2019}) to increase the preference for multiple stable branches, all else being equal.

While we stress that the statistical evidence in favor of \EOSs~that support multiple stable branches in the $M$-$R$ relation is weak, GW170817's preference for soft \EOSs, in conjunction with the existence of a 2 $M_{\odot}$ pulsar, could be interpreted as evidence for a strong phase transition between \rhonuc~and 2\rhonuc, although the precise onset density, pressure, and latent energy associated with such a phase transition are still largely uncertain.
Nonetheless, some theoretical studies of chiral effective field theory suggest that the purely hadronic model for the \EOS~will break down in this density range due to phase transitions~\cite{Tews2018a, Tews2018b, Tews2019, Furnstahl2015}.
This coincidence is intriguing, especially since none of our input candidate \EOSs~are computed within the chiral effective field theory framework.
It is therefore possible that we have observed exotic particles in the cores of coalescing \NSs~with GW170817.

%-------------------------------------------------
\section{Discussion}
\label{sec:disc}

We present a comprehensive nonparametric inference of the equation of state of neutron star matter, as informed by GW170817.
Nonparametric analyses allow for expansive model freedom, thereby mitigating the kind of systematic errors associated with parameterized \EOSs~while additionally providing transparent priors on physical quantities.
Our nonparametric approach does not assign any particular importance to specific densities or pressures \emph{a priori}, and the features observed in the inferred \EOS~\emph{a posteriori} are therefore driven by the data rather than prior beliefs.
We improve the nonparametric priors introduced in Landry \& Essick~\cite{Landry2019} by including additional tabulated \EOSs~from the literature and constructing separate priors for different underlying compositions.
We analyze publicly available data from GW170817~\cite{GW170817samples} after conditioning on the existence of massive pulsars~\cite{Antoniadis2013, Cromartie2019} and find that GW170817 favors soft \EOSs, with $p(2\rhonuc)=\PTwoRhonucAgnostic$ (\PTwoRhonucInformed) dyn/cm$^2$ and $p(6\rhonuc)=\PSixRhonucAgnostic$ (\PSixRhonucInformed) dyn/cm$^2$, in agreement with previous studies~\cite{GW170817eos, Landry2019}.
Given these constraints on the \EOS, we are able to infer properties of a canonical $1.4\,M_\odot$ neutron star.
In particular, we find \textit{a posteriori} medians and 90\% highest-probability-density credible regions of $\Lambda_{1.4}=\CanonicalLAgnostic$ (\CanonicalLInformed), $R_{1.4}=\CanonicalRAgnostic$ (\CanonicalRInformed) km, and a maximum mass $\Mmax = \MmaxAgnostic$ (\MmaxInformed) $M_\odot$ with our \emph{model-agnostic} (\emph{model-informed}) priors marginalized over composition.
We find mild evidence against a \BBH~progenitor, with $B^{\rm \NS}_{\BBH}=\BayesNSBHHispinAgnostic$, in accordance with analyses of individual tabulated \EOSs~\cite{GW170817modelselection} and the observation of electromagnetic counterparts~\cite{GW170817mma, GW170817grb}.
Intriguingly, and for the first time, we find a weak preference for \EOS~containing quark matter with our \emph{informed} prior.
We also find a preference for \EOSs~that support multiple stable branches in their $M$--$R$ and $M$--$\Lambda$ relations with our \emph{agnostic} prior when comparing results using \GW~and pulsar data against pulsar data alone.
While far from conclusive, we note that if the \EOS~supports multiple stable branches, there is a noticeable preference for dramatic softening around \rhonuc~followed by stiffening around 2\rhonuc, consistent with a phase transition and predictions for where chiral effective field theory may break down~\cite{Tews2018a, Tews2018b, Tews2019, Furnstahl2015}.
However, the exact onset density, pressure, and latent energy remain uncertain by at least a factor of a few.
We emphasize that these results are informed solely by GW170817 data and massive pulsar observations; our nonparametric prior processes were not constructed with any particular importance given to these densities, or to \EOSs~with particular phase transitions.

Using our posterior processes, we estimate the amount of baryonic mass dynamically ejected during the coalescence ($\Meject=\MejectAgnostic\,M_\odot$) and its velocity ($\Veject=\VejectAgnostic$), concluding that it must have been a subdominant component of the total ejected mass that powered the associated kilonova ($\gtrsim0.05\,M_\odot$~\cite{Coughlin2018}).

We also point out that the lower limit on the maximum spin frequency of \NSs~can approach the observed spin frequencies of some millisecond pulsars if they have relatively low masses (\result{$\lesssim M_\odot$}).
Because the masses of many millisecond pulsars are unknown, this may warrant a reexamination of the need for additional braking mechanisms, such as $r$-mode CFS instabilities.
Indeed, although the fastest-spinning known pulsar does not have a precise mass measurement, we find its observed spin is \result{$\gtrsim\Fbrk/2$} for all astrophysically plausible masses at 90\% confidence, and that it is consistent with \result{$f=\Fbrk$} at $>90\%$ confidence if it is lighter than $M_\odot$.

While our results already extend beyond previous studies, there are several ways our analysis might be further improved.
For example, one could incorporate an improved representation of the \GW~likelihood, perhaps through a \KDE~with better control over systematics associated with hard prior bounds, or explicitly account for uncertainty in the likelihood model due to the finite number of posterior samples available. However, it is believed that systematic uncertainties from such issues are currently dominated by statistical uncertainties.
Similarly, improved covariance kernels within our \GPs~could allow for even more model freedom beyond the already formidable range allowed by our mixture model over different hyperparameters.
In particular, kernels that support different length scales at different pressures, to further enhance our ability to model different levels of theoretical uncertainty at different pressures, could be promising.
Even with more complicated kernels, our hyperparameters will retain immediately interpretable meanings, such as how quickly the energy density can change with pressure.

Beyond these technical improvements, one could incorporate information from other observations in a more sophisticated way.
For example, instead of simply discarding any synthetic \EOSs~that do not support a $1.93\,M_\odot$ star based on the approximate 2-$\sigma$ lower limit from J0348+0432~\cite{Antoniadis2013} and J0740+6620~\cite{Cromartie2019}, we could instead use the entire uncertainty in such mass measurements to weight each \EOS, similar to what is proposed in Ref.~\cite{Miller2019}.
These weights can be combined with likelihoods from other \GW~observations as well as $M$--$R$ measurements from X-ray timing~\cite{Gendreau2012}, moment-of-inertia measurements from radio observations~\cite{Dewdney, Landry2018}, or even mass and frequency measurements from millisecond pulsars in a single self-consistent framework without the need for hard thresholds.
We will explore this avenue in future work.

Similarly, including better constraints on the theoretical uncertainty at low densities by matching to chiral effective field theory with theoretical uncertainties from truncating series expansions~\cite{Tews2018a, Tews2018b, Tews2019, Furnstahl2015} should further constrain our priors.
We note, though, that GW170817 produces posterior processes that already tend to follow the prior relatively closely below \rhonuc~(see Figure~\ref{fig:eos given n>1}).
Along the same lines, forcing our synthetic \EOSs~to asymptote to $c_s^2 \rightarrow c^2/3$ at high density, consistent with ultra-relativistic matter, may prove interesting.

Nonetheless, our nonparametric inference already provides novel results.
Because of the extreme model freedom supported in our prior processes, we see tantalizing hints of new physics and phase transitions within the cores of \NSs.
In particular, the suggestive preference for quark matter and \EOSs~that support multiple stable branches by factors of $\approx\BayesBranchesApprox$ imply the possible presence of a phase transition between \rhonuc~and 2\rhonuc~with no special significance given to these densities \textit{a priori}.
While the statistical evidence remains marginal at best, the agreement between these observations and predictions from theory could be taken as evidence that we have already seen new states of matter within \NS~cores.
Indeed, this demonstrates the key role \GWs~will play in determining the supranuclear \EOS~and the unique capabilities of nonparametric analyses.

In the near future, our nonparametric analyses will allow for further investigations into possible phase transitions in supranuclear matter, including the maximum sound-speed achieved within \NSs.
They provide for natural null tests of the universal-\EOS~hypothesis, but also allow for natural alternative hypotheses by spanning the space of \EOSs~allowed by causality and thermodynamic stability regardless of their underlying composition.
Finally, our analysis can naturally incorporate information from arbitrarily many sources, including constraints about high-density matter from the observation of massive pulsars and information about nuclear densities from coalescing \NSs~via \GWs~and accreting systems through X-ray timing, while self-consistently accounting for selection effects, astrophysical populations, and formation channels.
As demonstrated by our novel results, nonparametric analyses provide the best chance to capture new physics without systematic modeling errors using multi-messenger astrophysical observations and will only become more important in the years to come.

%-------------------------------------------------
\acknowledgments

The authors gratefully acknowledge their many useful discussions with Jocelyn Read, Maya Fishbach, Zoheyr Doctor, Amanda Farah, and Cole Miller, and also thank Armen Sedrakian for sharing several \EOS~tables.
This research was partially completed while at the Kavli Institute for Theoretical Physics, and was supported in part by the NSF under grant NSF PHY11-25915.
P.~L. is supported in part by the Natural Sciences and Engineering Research Council of Canada, and by NSF grants PHY 15-05124 and PHY 17-08081 to the University of Chicago.
R.~E. and D.~E.~H. are supported at the University of Chicago by the Kavli Institute for Cosmological Physics through an endowment from the Kavli Foundation and its founder Fred Kavli.
D.~E.~H. is also supported by NSF grant PHY-1708081, and gratefully acknowledges the Marion and Stuart Rice Award.
The authors also gratefully acknowledge the computational resources provided by the LIGO Laboratory and supported by NSF grants PHY-0757058 and PHY-0823459.

%-------------------------------------------------
\bibliography{gpr-eos-refs.bib}

%-------------------------------------------------
\appendix

\section{Details of nonparametric prior construction}
\label{sec:details of construction}

While Section~\ref{sec:inference} provides an overview of how we construct our priors, we report many important technical details here.
First, as in \LandryEssick, each tabulated $\varepsilon(p)$ is resampled to obtain a process for $\phi(p)$ using $K_\mathrm{se}$ and $K_\mathrm{wn}$ with hyperparameters optimized separately for each \EOS.
Because the data for each tabulated \EOS~represents a single function, we optimize hyperparameters using the marginal likelihood (see Section 5.4.1 of \cite{Rasmussen2006})
\begin{multline}\label{eqn:ml likelihood}
    \log P_\mathrm{ML}(\varepsilon|p,\vec{\sigma}) = -\frac{1}{2}(\varepsilon_i-\mu_i) \left(K^{-1}\right)_{ij} (\varepsilon_j-\mu_j) \\
        - \frac{N}{2}\log\left(2\pi\right) - \frac{1}{2}\log\left|K_{ij}\right|
\end{multline}
where $N$ is the dimensionality of $\varepsilon_i$ and $|K_{ij}|$ is the determinant of the covariance matrix.
This likelihood, which is the probability of obtaining the observed data given a \GP~with the specified hyperparameters, selects the best-fit element of the statistical model defined by our \GP.
We typically observe strong correlations between $\sigma$ and $l$, as increasing either increases the correlation between neighboring points.

We note that \EOSs~with sharp features, like strong first-order phase transitions, are somewhat difficult to model with a squared-exponential kernel because $K_\mathrm{se}$ strictly generates analytic functions.
Nonetheless, we find it is still possible to adequately represent the behavior seen in, e.g., tabulated \EOSs~that contain quark matter over the densities relevant for GW170817.

We then construct separate \GPs~to represent all tabulated \EOSs~belonging to the same composition and family using $K_\mathrm{se}$, $K_\mathrm{wn}$, and $K_\mathrm{mv}$.
Because these \GPs~are meant to emulate the typical behavior of a group of proposed \EOSs~rather than reproduce a single function, we select hyperparameters based on a cross-validation likelihood (see Section 5.4.2 of \cite{Rasmussen2006} and Eq.~(\ref{eqn:cv likelihood})).
The cross-validation likelihood selects hyperparameters that produce synthetic \EOSs~with a variance similar to what is seen between elements of the training set.
We select a single set of optimal hyperparameters based on the cross-validation likelihood separately for each composition and family.

Similar to $P_\mathrm{ML}$, $P_\mathrm{CV}$ shows strong correlations between $\sigma$ and $l$.
These are often independent of $\sigma_\mathrm{obs}$ and $m$, which themselves show strong degeneracies.
Usually, the input \EOSs~strongly prefer a specific ($\sigma$, $l$) pair, indicative of the typical correlations and length scales in the underlying data.
Additionally, the cross-validation requires at least some spread in the conditioned processes, meaning that as long as either $\sigma_\mathrm{obs}$ or $m$ is sufficiently large, there is little preference for the precise combination.

After obtaining \GPs~for each composition and family combination, we construct \emph{agnostic} and \emph{informed} \GPs~for each composition, both of which are conditioned on the \GPs~for all families within that composition.
Each family is equally weighted so that those containing many slightly different tabulated \EOSs~are not more strongly weighted than families with fewer \EOSs, as the number of tabulated \EOSs~in each family may depend on different authors' propensities to publish.
We note that this is not a unique choice and different relative weights would produce different priors.

As described in Section~\ref{sec:setup}, we again use the cross-validation likelihood (Eq~\ref{eqn:cv likelihood}) with $K_\mathrm{se}$, $K_\mathrm{wn}$, and $K_\mathrm{mv}$ to obtain processes that emulate the behavior seen within each composition, observing similar correlations to those observed within each combination of composition and family.
However, instead of selecting a single set of hyperparameters, we sample from $P_\mathrm{CV}$ as in Eq.~\ref{eqn:gp mixture model}.
This generates a mixture model of many different \GPs~from which we sample when drawing from our priors.

We generate \emph{informed} priors via simulated annealing, repeatedly Monte-Carlo sampling from the hyperprior as we slowly decrease the temperature in order to find the high-likelihood portions of hyperparameter space.
However, we note that taking the limit $T\rightarrow\infty$ with arbitrary prior bounds may not produce the range of variability desired for the \emph{agnostic} prior.
Specifically, we identify several regions of hyperparameter space that produce nearly identical conditioned processes, several of which we consider ``too tight'' as they do not produce a broad range of synthetic \EOSs.
Therefore, we impose several additional hyperprior constraints for the \emph{agnostic} priors.
We require $\sigma\geq1$, $\sigma_\mathrm{obs}\geq1$, $m\geq1$, and $(\sigma_\mathrm{obs}^2+m^2)\geq 2\sigma^2$.
The first condition allows the resulting processes to deviate significantly from the mean, while the second and third conditions make the conditioned process less sensitive to the specific behavior seen in the tabulated \EOSs.
The final condition on the ratio of hyperparameters avoids situations where the modeling uncertainty ($\sigma_\mathrm{obs}$, $m$) is significantly smaller than the marginal $K_\mathrm{se}$ uncertainty ($\sigma$).
When that happens, the conditioned process returns a weighted average of the input \EOSs~with a variance similar to the modeling uncertainty rather than $\sigma$, as would be expected when taking the average of many independent Gaussian-distributed variates.
In the opposite extreme, where the modeling uncertainty is much larger than $\sigma$, the resulting conditioned process follows the prior mean with variances characteristic of $\sigma$, which produce reasonably broad synthetic \EOSs~because we require $\sigma\geq1$.
We additionally allow $l$ to vary over nearly an order of magnitude.

%-------------------------------------------------

\section{Optimal Gaussian kernel density estimate representations of the gravitational-wave likelihood}
\label{sec:KDE}

Our Gaussian \KDE~model for the \GW~likelihood is constructed in the four-dimensional space spanned by $M_1$, $M_2$, $\Lambda_1$, and $\Lambda_2$.
We assume a diagonal covariance ($C_{ij}$) within our Gaussian kernels, optimizing the bandwidths directly by varying a scale parameter for each dimension's sample variance so that
\begin{equation}
    C_{ij} = b^2 \sigma_i^2 \delta_{ij}\quad \left| \quad \sigma^2_i = \frac{1}{N}\sum\limits_\alpha \left(x^{(\alpha)}_i\right)^2 - \left(\frac{1}{N}\sum\limits_\alpha x^{(\alpha)}_i\right)^2\right.
\end{equation}
Pragmatically, this is done by whitening the samples with the sample variance in each dimension separately and then optimizing a scale parameter for a covariance proportional to the identity matrix.
This is achieved by directly maximizing a leave-one-out cross-validation likelihood as a function of the scale parameter $b$
\begin{equation}
    \log L_\mathrm{CV} = \sum\limits_i^N \log\left( \frac{1}{N-1}\sum\limits_{j\neq i}^{N-1} k(x_i, x_j; b)\right)
\end{equation}
where $k$ is our multi-dimensional Gaussian kernel.
Using the available samples~\cite{GW170817samples}, we find optimal bandwidths of $b_\mathrm{opt}=0.1247$ ($0.0905$) for the low-spin (high-spin) data set.

Furthermore, to account for hard prior bounds like $\Lambda_{1,2}\geq0$, we reflect our samples across such boundaries, \textit{de facto} forcing the \KDE's derivative to vanish at the boundary in directions perpendicular to the boundary.
This can introduce systematic biases beyond those produced by the smoothing inherent in all \KDE~models.
We measure the scale of their impact on Monte-Carlo integrals conducted along the boundary by comparing estimates with and without reflected samples.
Note that the relative weight assigned to samples far from the boundary, i.e. most of the \BNS~Monte-Carlo points, are virtually unaffected by such issues, and these primarily concern \NSBH, \BHNS, and \BBH~models.

%-------------------------------------------------

\section{Tabulated equations of state used to condition nonparametric priors}
\label{sec:tabulated eos}

Table~\ref{table:tabulated eos} lists the tabulated candidate \EOSs~used to condition our nonparametric priors, including their \Mmax~and source references.

\begin{table*}
    \caption{
        Tabulated \EOSs~grouped in the same way we construct our priors, first by combining members of each family of underlying nuclear effective forces separately, and then combining separate families for each composition.
    }
    \label{table:tabulated eos}
    \begin{minipage}{0.49\textwidth}
    \begin{tabular}{ccccl}
    \hline
    Composition                & Family               & Moniker          & $M_\mathrm{max}\ [M_\odot]$ & Reference \\
    \hline
    \hline
    \multirow{33}{*}{Hadronic} & \multirow{7}{*}{BSK} & \texttt{bsk20}   &  $2.16$ & \multirow{2}{*}{\cite{Goriely2010}} \\
                               &                      & \texttt{bsk21}   &  $2.27$ & \\
                                                      \cline{3-5}
                               &                      & \texttt{bsk22}   &  $2.26$ & \multirow{5}{*}{\cite{Goriely2013}} \\
                               &                      & \texttt{bsk23}   &  $2.27$ & \\
                               &                      & \texttt{bsk24}   &  $2.28$ & \\
                               &                      & \texttt{bsk25}   &  $2.22$ & \\
                               &                      & \texttt{bsk26}   &  $2.15$ & \\
                               \cline{2-5}
                               & \multirow{2}{*}{BSR} & \texttt{bsr2}    &  $2.38$ & \multirow{2}{*}{\cite{Agarwal2010}} \\
                               &                      & \texttt{bsr6}    &  $2.43$ & \\
                               \cline{2-5}
                               & \multirow{3}{*}{DD}  & \texttt{dd2}     &  $2.42$ & \multirow{1}{*}{\cite{Banik2014}} \\
                               &                      & \texttt{ddhd}    &  $2.14$ & \multirow{1}{*}{\cite{Gaitanos2004}} \\
                               &                      & \texttt{ddme2}   &  $2.48$ & \multirow{1}{*}{\cite{Lalazissis2005}} \\
                               \cline{2-5}
                               & \multirow{1}{*}{ENG} & \texttt{eng}     &  $2.24$ & \multirow{1}{*}{\cite{Engvik1995}} \\
                               \cline{2-5}
                               & \multirow{1}{*}{GM}  & \texttt{gm1}     &  $2.36$ & \multirow{1}{*}{\cite{Glendenning1991}} \\
                               \cline{2-5}
                               & \multirow{2}{*}{KDE} & \texttt{kde0v}   &  $1.96$ & \multirow{1}{*}{\cite{Gulminelli2015}} \\
                               &                      & \texttt{kde0v1}  &  $1.97$ & \multirow{1}{*}{\cite{Agarwal2005}} \\
                               \cline{2-5}
                               & \multirow{1}{*}{MPA} & \texttt{mpa1}    &  $2.46$ & \multirow{1}{*}{\cite{Muther1987}} \\
                               \cline{2-5}
                               & \multirow{1}{*}{NL}  & \texttt{nl3}     &  $2.77$ & \multirow{1}{*}{\cite{Lalazissis1997}} \\
                               \cline{2-5}
                               & \multirow{1}{*}{R}   & \texttt{rs}      &  $2.12$ & \multirow{1}{*}{\cite{Friedrich1986}} \\
                               \cline{2-5}
                               & \multirow{9}{*}{SK}  & \texttt{sk255}   &  $2.14$ & \multirow{2}{*}{\cite{Agarwal2003}} \\
                               &                      & \texttt{sk272}   &  $2.23$ & \\
                                                      \cline{3-5}
                               &                      & \texttt{ski2}    &  $2.16$ & \multirow{4}{*}{\cite{Reinhard1995}} \\
                               &                      & \texttt{ski3}    &  $2.24$ & \\
                               &                      & \texttt{ski4}    &  $2.17$ & \\
                               &                      & \texttt{ski5}    &  $2.24$ & \\
                                                      \cline{3-5}
                               &                      & \texttt{ski6}    &  $2.19$ & \multirow{1}{*}{\cite{Nazarewicz1996}} \\
                               &                      & \texttt{skmp}    &  $2.11$ & \multirow{1}{*}{\cite{Bennour1989}} \\
                               &                      & \texttt{skop}    &  $1.97$ & \multirow{1}{*}{\cite{Reinhard1999}} \\
                               \cline{2-5}
                               & \multirow{4}{*}{SLY} & \texttt{sly230a} &  $2.10$ & \multirow{1}{*}{\cite{Chabanat1997}} \\
                                                      \cline{3-5}
                               &                      & \texttt{sly2}    &  $2.05$ & \multirow{2}{*}{\cite{Chabanat1995}} \\
                               &                      & \texttt{sly9}    &  $2.05$ & \\
                                                      \cline{3-5}
                               &                      & \texttt{sly}     &  $2.16$ & \multirow{1}{*}{\cite{Douchin2001}} \\
                               \cline{2-5}
                               & \multirow{1}{*}{TM}  & \texttt{tm1}     &  $2.18$ & \multirow{1}{*}{\cite{Sugahara1994}} \\
    \hline
    \end{tabular}
    \end{minipage}
    \begin{minipage}{0.49\textwidth}
    \begin{tabular}{ccccl}
    \hline
    Composition                & Family               & Moniker          & $M_\mathrm{max}\ [M_\odot]$ & Reference \\
    \hline
    \hline
    \multirow{9}{*}{Hyperonic} & \multirow{2}{*}{BSR} & \texttt{bsr2y}   &  $2.00$ & \multirow{2}{*}{\cite{Fortin2016}} \\
                               &                      & \texttt{bsr6y}   &  $2.02$ & \\
                               \cline{2-5}
                               & \multirow{2}{*}{DD}  & \texttt{dd2y}    &  $2.00$ & \multirow{2}{*}{\cite{Fortin2016}} \\
                               &                      & \texttt{ddme2y}  &  $2.09$ & \\
                               \cline{2-5}
                               & \multirow{2}{*}{GM}  & \texttt{gm1b}    &  $1.99$ & \multirow{1}{*}{\cite{Gusakov2014}} \\
                               &                      & \texttt{gm1y}    &  $2.02$ & \multirow{1}{*}{\cite{Fortin2016}} \\
                               \cline{2-5}
                               & \multirow{1}{*}{H}   & \texttt{h4}      &  $2.03$ & \multirow{1}{*}{\cite{Lackey2006}} \\
                               \cline{2-5}
                               & \multirow{1}{*}{NL}  & \texttt{nl3y}    &  $2.31$ & \multirow{1}{*}{\cite{Fortin2016}} \\
                               \cline{2-5}
                               & \multirow{1}{*}{TM}  & \texttt{tm1c}    &  $2.06$ & \multirow{1}{*}{\cite{Gusakov2014}} \\
    \hline
    \multirow{8}{*}{Quark}     & \multirow{2}{*}{ALF} & \texttt{alf2}    &  $2.09$ & \multirow{2}{*}{\cite{Alford2005}} \\
                               &                      & \texttt{alf4}    &  $1.94$ & \\
                               \cline{2-5}
                               & \multirow{4}{*}{DDQ} & \texttt{ddq0625}   & $1.93$ & \multirow{4}{*}{\cite{Bonanno2011}} \\
                               &                      & \texttt{ddq0630}   & $2.07$ & \\
                               &                      & \texttt{ddq0825}   & $1.99$ & \\
                               &                      & \texttt{ddq0830}   & $2.08$ & \\
                               \cline{2-5}
                               & \multirow{2}{*}{HQC} & \texttt{hqc18}   &  $2.05$ & \multirow{2}{*}{\cite{Baym2019}} \\
                               &                      & \texttt{hqc19}   &  $2.07$ & \\
    \hline
    \end{tabular}
    \vspace{7.22cm}
    \end{minipage}
\end{table*}

%-------------------------------------------------

\section{Supplementary figures and tables}
\label{sec:supplementary}

\begin{figure*}
    \begin{center}
        \emph{model-informed} \\
    \end{center}
    \begin{minipage}{0.45\textwidth}
        \begin{center}
            \includegraphics[width=1.0\textwidth, clip=True, trim=0.0cm 1.2cm 0.0cm 0.50cm]{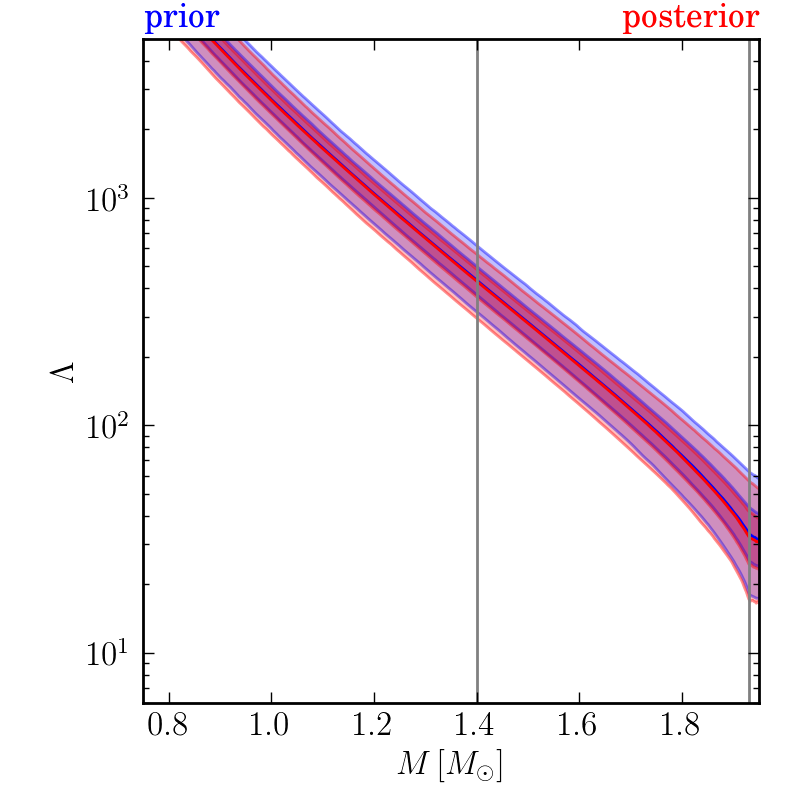} \\
            \includegraphics[width=1.0\textwidth, clip=True, trim=0.0cm 1.2cm 0.0cm 0.45cm]{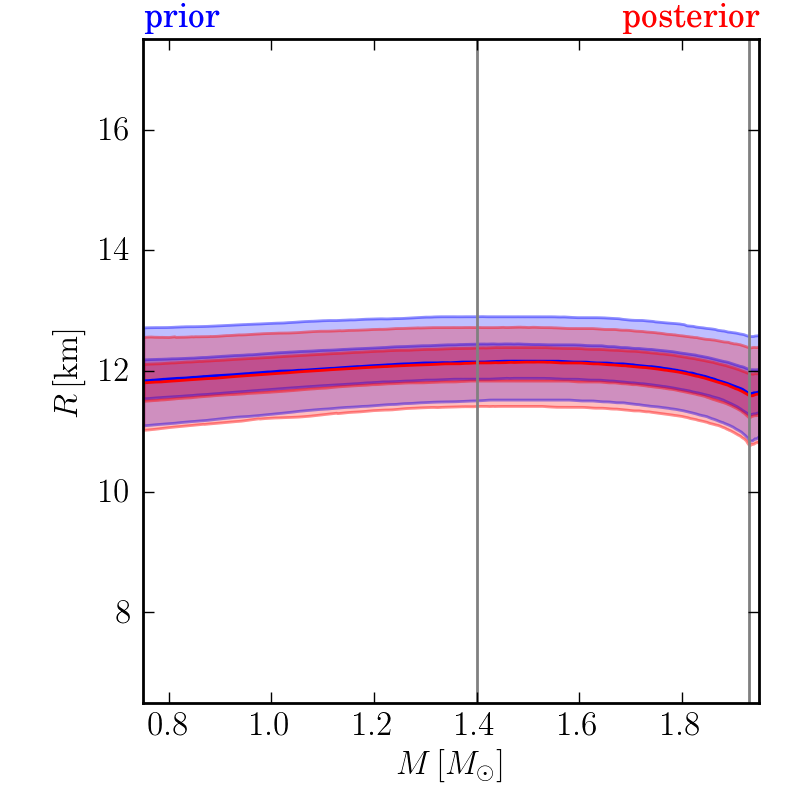} \\
            \includegraphics[width=1.0\textwidth, clip=True, trim=0.0cm 0.0cm 0.0cm 0.45cm]{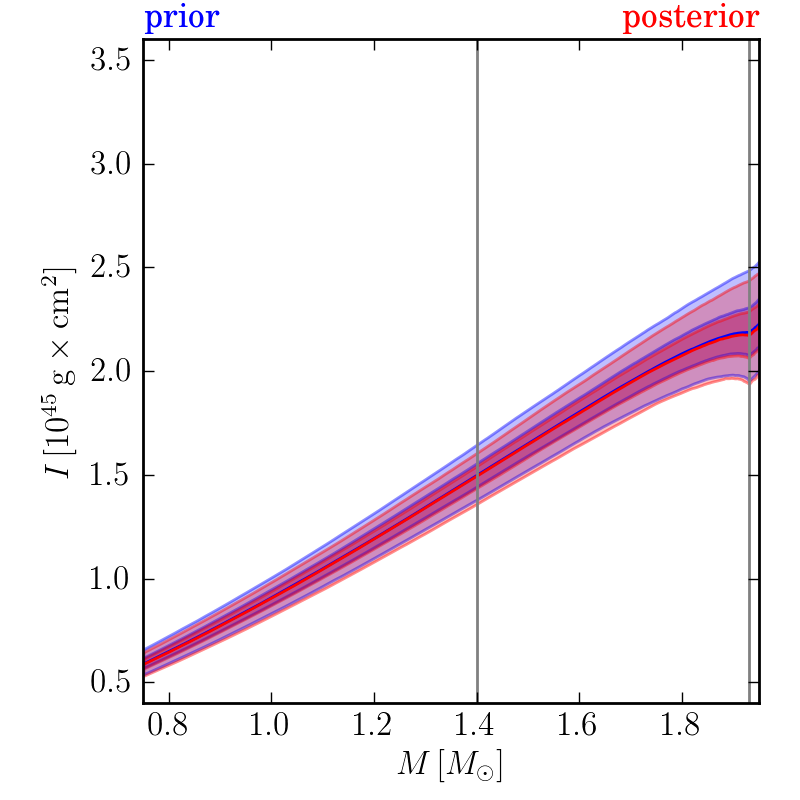}
        \end{center}
    \end{minipage}
    \begin{minipage}{0.45\textwidth}
        \begin{center}
            \includegraphics[width=1.0\textwidth, clip=True, trim=0.0cm 1.2cm 0.0cm 0.50cm]{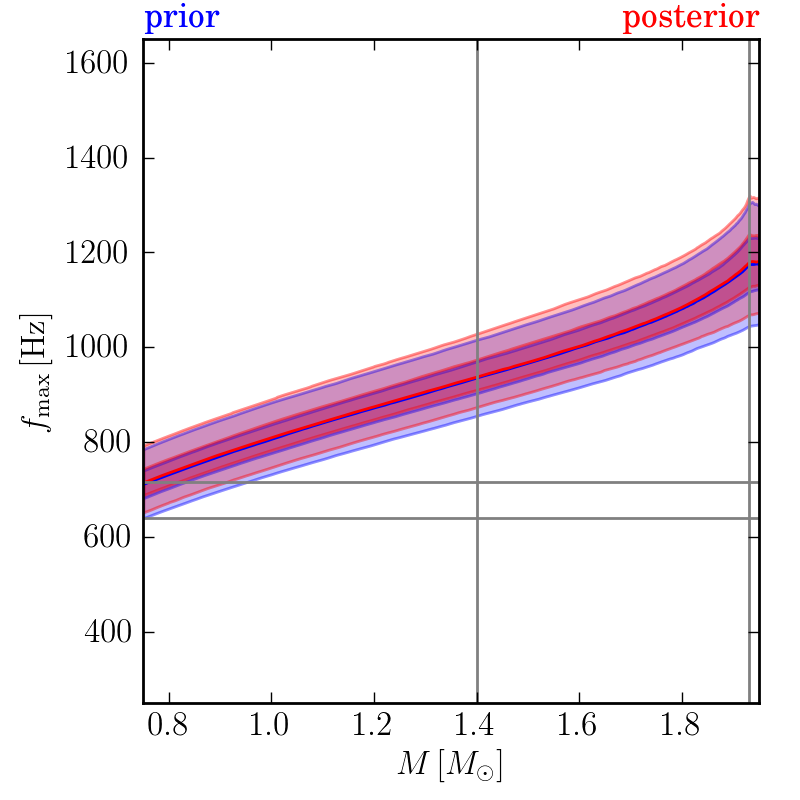} \\
            \includegraphics[width=1.0\textwidth, clip=True, trim=0.0cm 1.2cm 0.0cm 0.45cm]{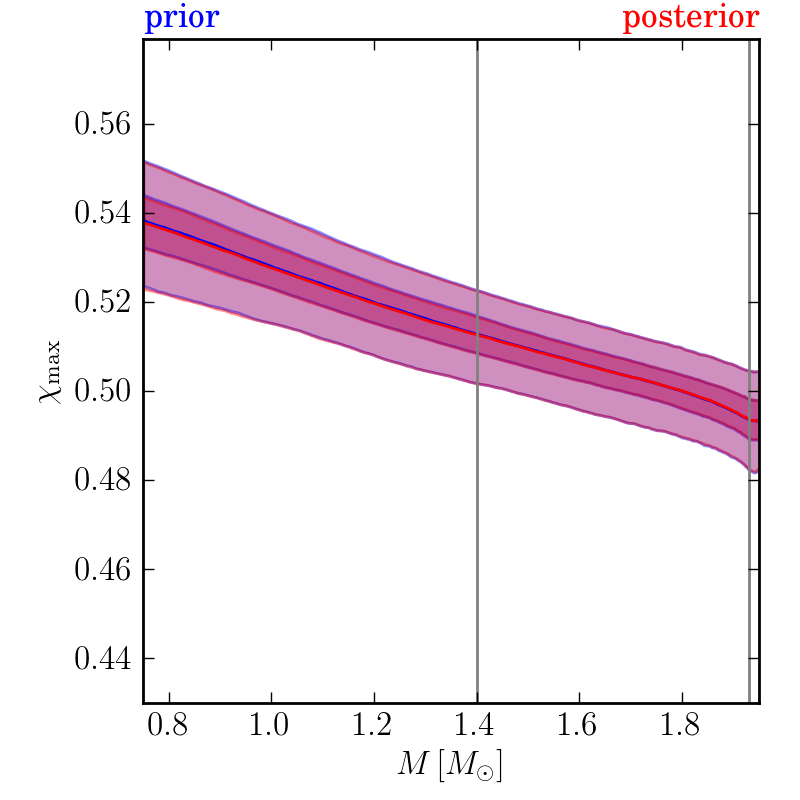} \\
            \includegraphics[width=1.0\textwidth, clip=True, trim=0.0cm 0.0cm 0.0cm 0.45cm]{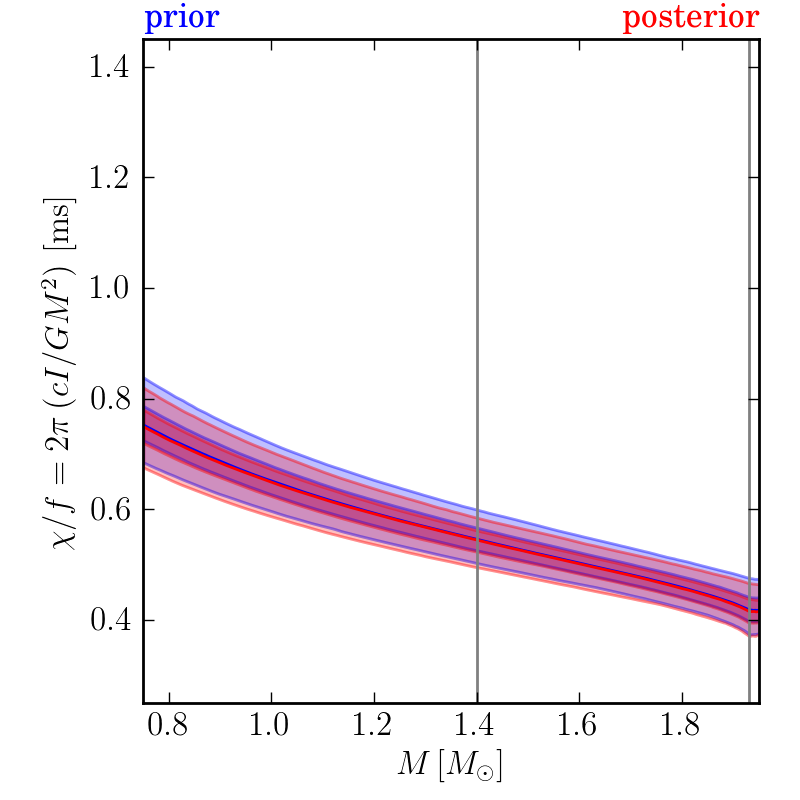}
        \end{center}
    \end{minipage}
    \caption{
        Processes relating a few macroscopic observables after marginalizing over \EOS-composition with the \emph{informed} prior.
        Prior (\emph{cyan}) and posterior (\emph{magenta}) processes for (\emph{left}) tidal deformability ($\Lambda$; \emph{top}), radius ($R$; \emph{middle}), and moment-of-inertia ($I$; \emph{bottom}) as functions of mass as well as (\emph{right}) the maximum spin frequency (\Fbrk; \emph{top}), maximum dimensionless spin parameter (\Chibrk; \emph{middle}), and $\chi/f$ (\emph{bottom}).
        Shaded regions denote the 50\% and 90\% symmetric credible regions for the marginal distribution of each observable at each mass.
        Solid lines denote the median and vertical lines denote canonical $1.4\,M_\odot$ stars.
        Horizontal grey lines in the \emph{top}-\emph{right} panel denote the measured spin frequencies of J1748--2446ad (\externalresult{$716\,\mathrm{Hz}$}~\cite{Hessels2006}) and B1937+241 (\externalresult{641 Hz}~\cite{Backer1982}), which lie below but near the lower limits for \Fbrk.
    }
    \label{fig:process marginf}
\end{figure*}

\begin{figure*}
    \begin{minipage}{0.49\textwidth}
        \begin{center}
            \emph{model-agnostic} \\
            \includegraphics[width=1.0\textwidth, clip=True, trim=0.5cm 1.9cm 0.5cm 0.4cm]{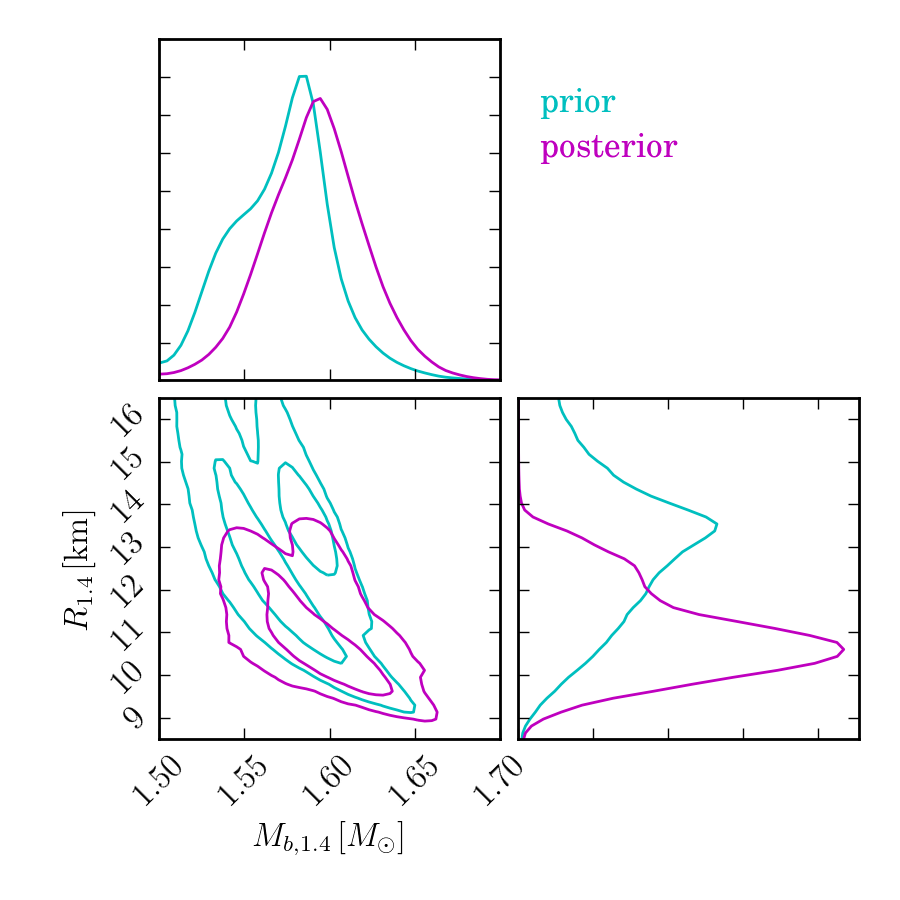} \\
            \includegraphics[width=1.0\textwidth, clip=True, trim=0.5cm 1.9cm 0.5cm 5.0cm]{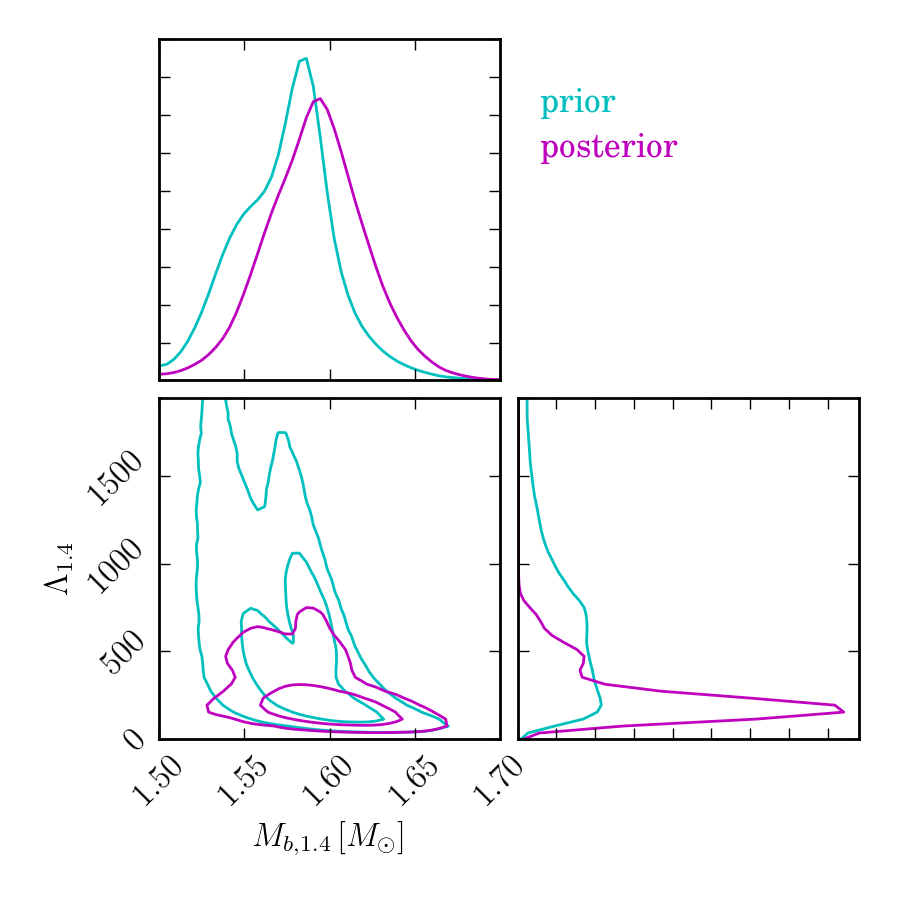} \\
            \includegraphics[width=1.0\textwidth, clip=True, trim=0.5cm 0.6cm 0.5cm 5.0cm]{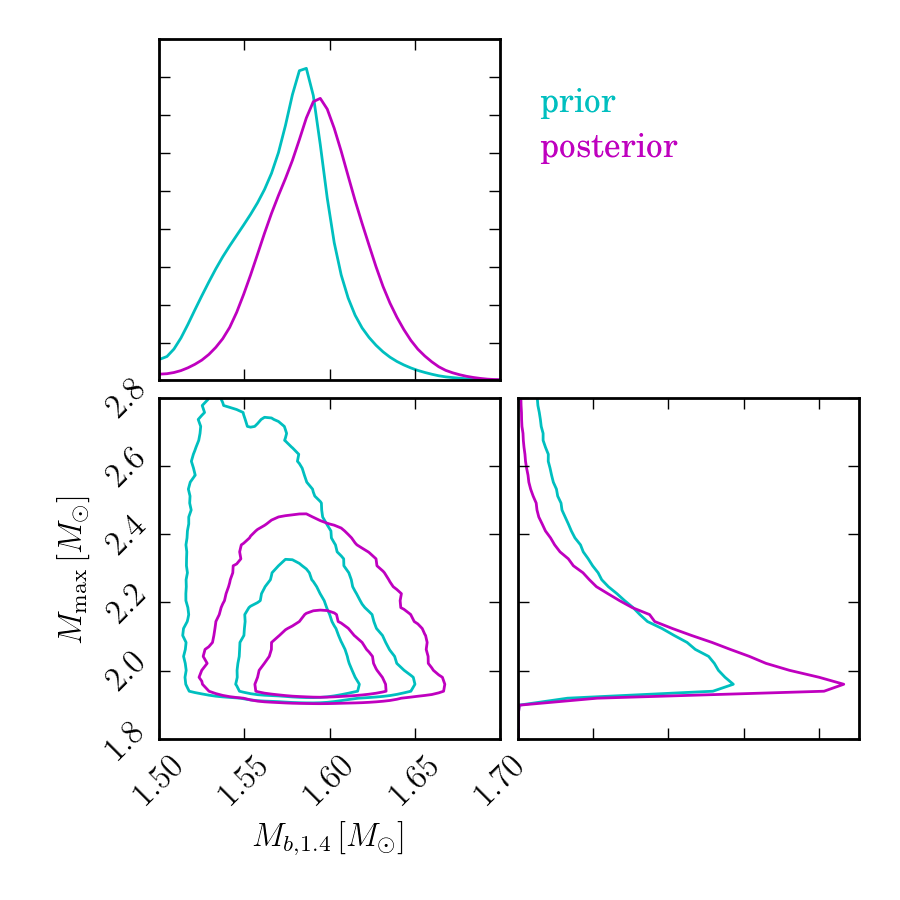}
        \end{center}
    \end{minipage}
    \begin{minipage}{0.49\textwidth}
        \begin{center}
            \emph{model-informed} \\
            \includegraphics[width=1.0\textwidth, clip=True, trim=0.5cm 1.9cm 0.5cm 0.4cm]{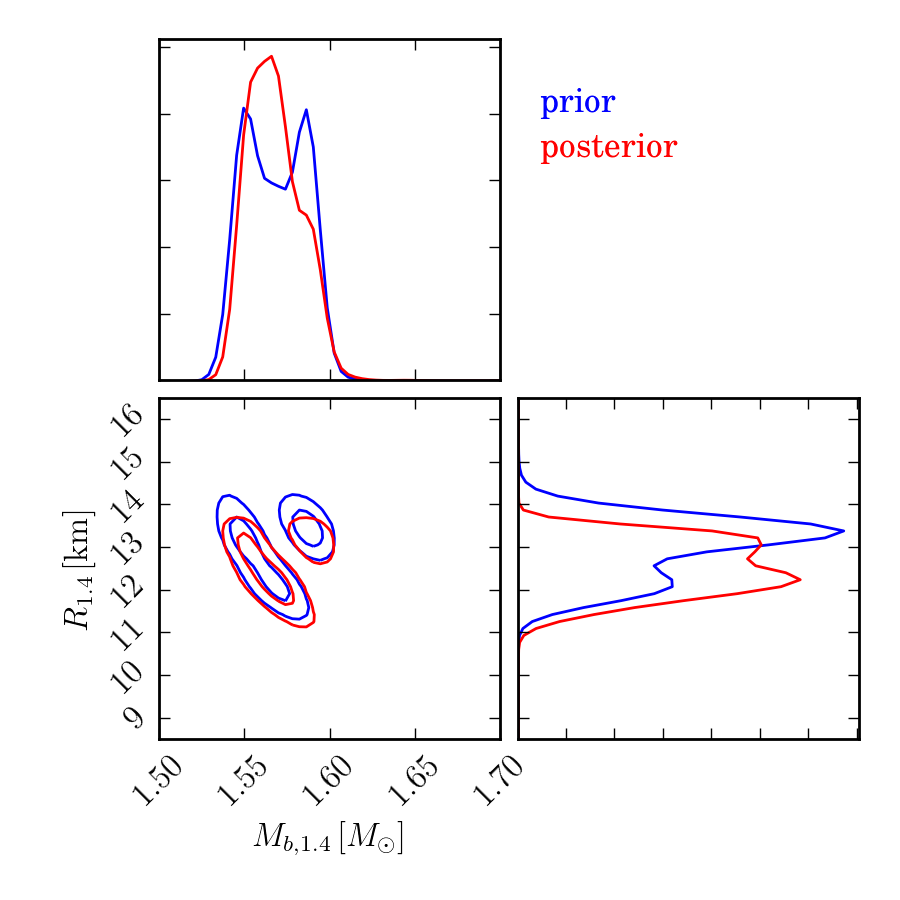} \\
            \includegraphics[width=1.0\textwidth, clip=True, trim=0.5cm 1.9cm 0.5cm 5.0cm]{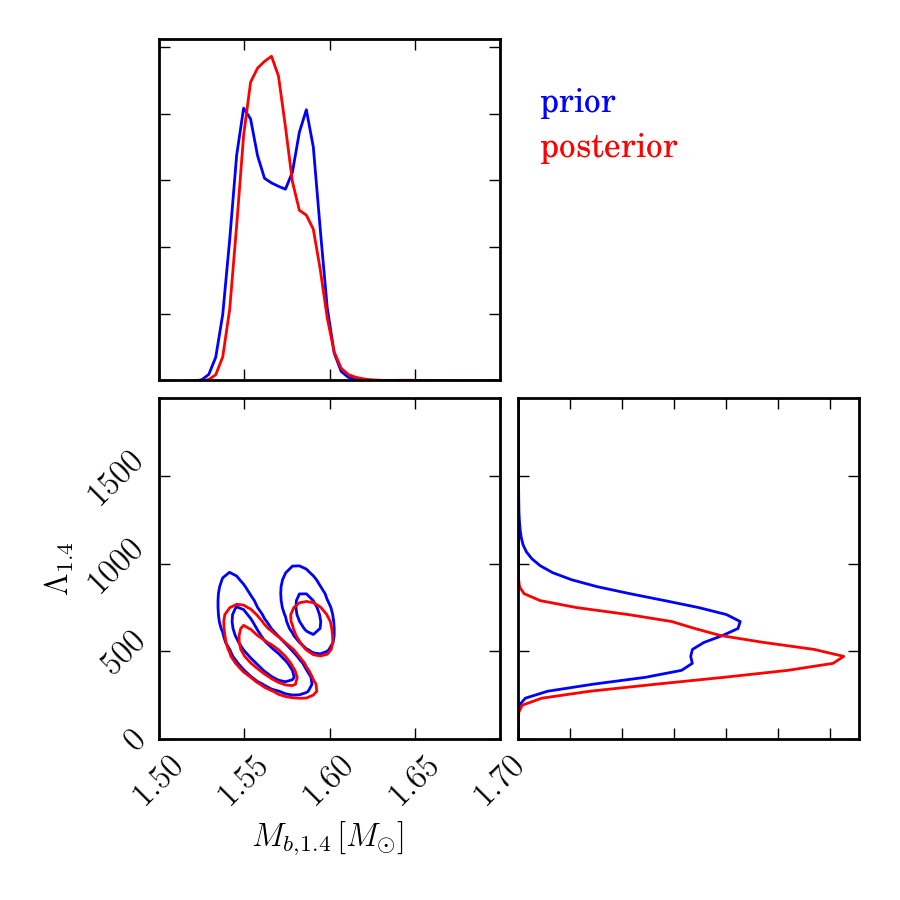} \\
            \includegraphics[width=1.0\textwidth, clip=True, trim=0.5cm 0.6cm 0.5cm 5.0cm]{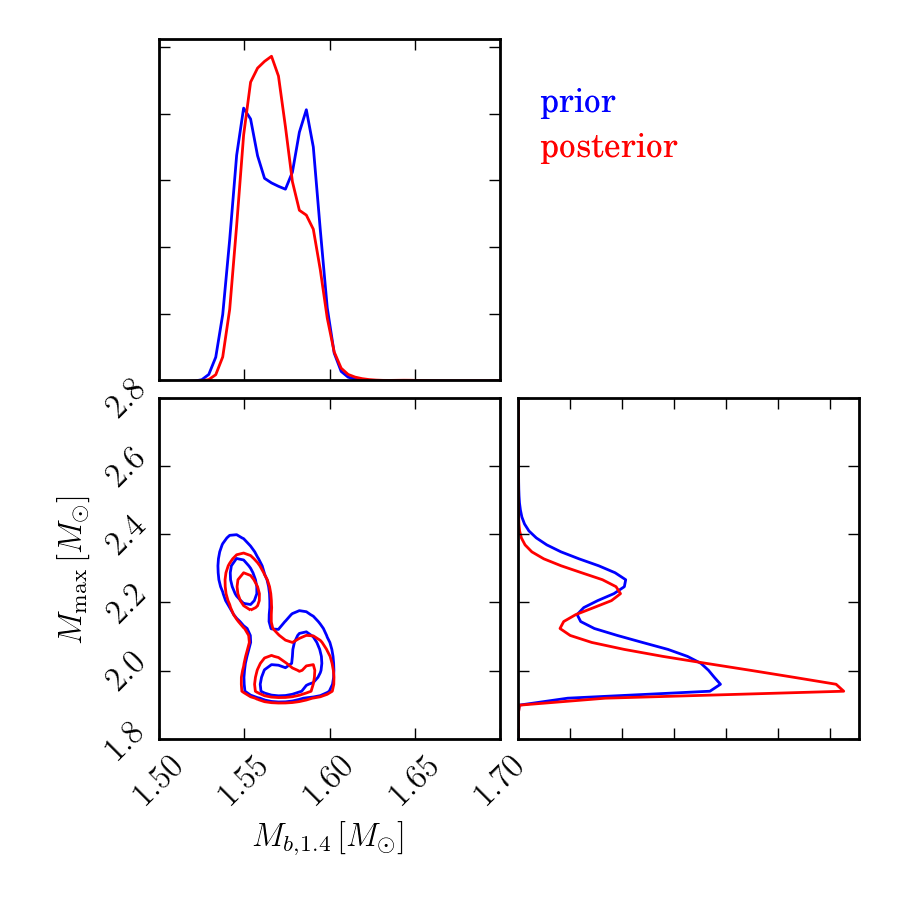}
        \end{center}
    \end{minipage}
    \caption{
        Distributions for $M_{b,1.4}$, $M_\mathrm{max}$, $\Lambda_{1.4}$, and $R_{1.4}$ after marginalizing over \EOS-composition.
        (\emph{Left}) \emph{model-agnostic} prior (\emph{cyan}), posterior (\emph{magenta}), and low-spin marginal-likelihood (\emph{green}).
        (\emph{Right}) \emph{model-informed} prior (\emph{blue}), posterior (\emph{red}), and low-spin marginal-likelihood (\emph{green}).
        Contours in the joint distributions denote minimal 50\% and 90\% credible regions.
        The bimodal behavior seen with the \emph{agnostic} posterior is a combination of the inherited multimodal likelihood from GW170817 (see Figure 11 of Ref.~\cite{GW170817properties}) as well as the preference for multiple stable branches.
        \EOSs~with multiple stable branches \emph{only} inhabit the low-$R_\mathrm{1.4}$ mode, for example, while \EOSs~with a single stable branch inhabit both.
        The multimodal behavior seen with the \emph{informed} prior is mostly due to the tight constraints imposed for each composition separately, which result in the separate peaks observed in these canonical observables.
    }
    \label{fig:universal marg}
\end{figure*}

We present a few additional tables and figures relevant for our analyses. Tables~\ref{table:macroscopic by composition}-\ref{table:ilov central by composition} contain results broken down by composition. Fig.~\ref{fig:process marginf} shows \emph{informed}-prior results for functional relations between generic \NS~observables, and Fig.~\ref{fig:universal marg} plots distributions for canonical and maximum-mass quantities.

\begin{table*}
    \caption{
        Medians \textit{a posteriori} and highest-probability-density 90\% credible regions for macroscopic observables associated with GW170817, with priors broken down according to the composition of the \EOS~upon which they were conditioned.
    }
    \label{table:macroscopic by composition}
    \begin{tabular}{ccrrrrrrrcc}
        \hline
        \multicolumn{2}{c}{Prior ($\mathcal{H}_i$)}     & \multicolumn{1}{c}{$M_1\,[M_\odot]$} & \multicolumn{1}{c}{$M_2\,[M_\odot]$} & \multicolumn{1}{c}{$\Lambda_1$} & \multicolumn{1}{c}{$\Lambda_2$} & \multicolumn{1}{c}{$\tilde\Lambda$} & \multicolumn{1}{c}{$R_1\,[\mathrm{km}]$} & \multicolumn{1}{c}{$R_2\,[\mathrm{km}]$} & \multicolumn{1}{c}{$\rho_{c,1}\,[10^{14}\,\mathrm{g}/\mathrm{cm}^3]$} & \multicolumn{1}{c}{$\rho_{c,2}\,[10^{14}\,\mathrm{g}/\mathrm{cm}^3]$} \\
        \hline
        \hline
        \multirow{4}{*}{\emph{informed}} & Hadronic     & \MOneHadInf & \MTwoHadInf & \LOneHadInf & \LTwoHadInf & \LTildeHadInf & \ROneHadInf & \RTwoHadInf & \RhocOneHadInf & \RhocTwoHadInf \\
                                         & Hyperonic    & \MOneHypInf & \MTwoHypInf & \LOneHypInf & \LTwoHypInf & \LTildeHypInf & \ROneHypInf & \RTwoHypInf & \RhocOneHypInf & \RhocTwoHypInf \\
                                         & Quark        & \MOneQrkInf & \MTwoQrkInf & \LOneQrkInf & \LTwoQrkInf & \LTildeQrkInf & \ROneQrkInf & \RTwoQrkInf & \RhocOneQrkInf & \RhocTwoQrkInf \\
        \cline{2-11}
        \multirow{4}{*}{\emph{agnostic}} & Hadronic     & \MOneHadAgn & \MTwoHadAgn & \LOneHadAgn & \LTwoHadAgn & \LTildeHadAgn & \ROneHadAgn & \RTwoHadAgn & \RhocOneHadAgn & \RhocTwoHadAgn \\
                                         & Hyperonic    & \MOneHypAgn & \MTwoHypAgn & \LOneHypAgn & \LTwoHypAgn & \LTildeHypAgn & \ROneHypAgn & \RTwoHypAgn & \RhocOneHypAgn & \RhocTwoHypAgn \\
                                         & Quark        & \MOneQrkAgn & \MTwoQrkAgn & \LOneQrkAgn & \LTwoQrkAgn & \LTildeQrkAgn & \ROneQrkAgn & \RTwoQrkAgn & \RhocOneQrkAgn & \RhocTwoQrkAgn \\
        \hline
    \end{tabular}
\end{table*}

\begin{table*}
    \caption{
        Medians \textit{a posteriori} and highest-probability-density 90\% credible regions for a few canonical macroscopic quantities, with priors broken down according to the composition of the \EOS~upon which they were conditioned.
    }
    \label{table:ilov by composition}
    \begin{tabular}{ccrrcrr}
        \hline
        \multicolumn{2}{c}{Prior ($\mathcal{H}_i$)}     & \multicolumn{1}{c}{$\Lambda_{1.4}$} & \multicolumn{1}{c}{$R_{1.4}\,[\mathrm{km}]$} & \multicolumn{1}{c}{$I_{1.4}\,[10^{45}\,\mathrm{g}\,\mathrm{cm}^2]$} & \multicolumn{1}{c}{$M_{b,1.4}\,[M_\odot]$} & \multicolumn{1}{c}{$\Mmax\,[M_\odot]$} \\
        \hline
        \hline
        \multirow{4}{*}{\emph{informed}} & Hadronic     & \CanonicalLHadInf & \CanonicalRHadInf & \CanonicalIHadInf & \CanonicalMbHadInf & \MmaxHadInf \\
                                         & Hyperonic    & \CanonicalLHypInf & \CanonicalRHypInf & \CanonicalIHypInf & \CanonicalMbHypInf & \MmaxHypInf \\
                                         & Quark        & \CanonicalLQrkInf & \CanonicalRQrkInf & \CanonicalIQrkInf & \CanonicalMbQrkInf & \MmaxQrkInf \\
        \cline{2-7}
        \multirow{4}{*}{\emph{agnostic}} & Hadronic     & \CanonicalLHadAgn & \CanonicalRHadAgn & \CanonicalIHadAgn & \CanonicalMbHadAgn & \MmaxHadAgn \\
                                         & Hyperonic    & \CanonicalLHypAgn & \CanonicalRHypAgn & \CanonicalIHypAgn & \CanonicalMbHypAgn & \MmaxHypAgn \\
                                         & Quark        & \CanonicalLQrkAgn & \CanonicalRQrkAgn & \CanonicalIQrkAgn & \CanonicalMbQrkAgn & \MmaxQrkAgn \\
        \hline
    \end{tabular}
\end{table*}

\begin{table*}
    \caption{
        Medians \textit{a posteriori} and highest-probability-density 90\% credible regions for canonical central densities and pressures at reference densities, with priors broken down according to the composition of the \EOS~upon which they were conditioned.
    }
    \label{table:ilov central by composition}
    \begin{tabular}{cccccc}
        \hline
        \multicolumn{2}{c}{Prior ($\mathcal{H}_i$)}     & \multicolumn{1}{c}{$\rho_{c,1.4}\,[\mathrm{g}/\mathrm{cm}^3]$} & \multicolumn{1}{c}{$p(\rhonuc)\,[\mathrm{dyn}/\mathrm{cm}^2]$} & \multicolumn{1}{c}{$p(2\rhonuc)\,[\mathrm{dyn}/\mathrm{cm}^2]$} & \multicolumn{1}{c}{$p(6\rhonuc)\,[\mathrm{dyn}/\mathrm{cm}^2]$} \\
        \hline
        \hline
        \multirow{4}{*}{\emph{informed}} & Hadronic     & \CanonicalRhocHadInf & \POneRhonucHadInf & \PTwoRhonucHadInf & \PSixRhonucHadInf \\
                                         & Hyperonic    & \CanonicalRhocHypInf & \POneRhonucHypInf & \PTwoRhonucHypInf & \PSixRhonucHypInf \\
                                         & Quark        & \CanonicalRhocQrkInf & \POneRhonucQrkInf & \PTwoRhonucQrkInf & \PSixRhonucQrkInf \\
        \cline{2-6}
        \multirow{4}{*}{\emph{agnostic}} & Hadronic     & \CanonicalRhocHadAgn & \POneRhonucHadAgn & \PTwoRhonucHadAgn & \PSixRhonucHadAgn \\
                                         & Hyperonic    & \CanonicalRhocHypAgn & \POneRhonucHypAgn & \PTwoRhonucHypAgn & \PSixRhonucHypAgn \\
                                         & Quark        & \CanonicalRhocQrkAgn & \POneRhonucQrkAgn & \PTwoRhonucQrkAgn & \PSixRhonucQrkAgn \\
        \hline
    \end{tabular}
\end{table*}

%------------------------

%-------------------------------------------------
\end{document}